\documentclass[12pt]{article}
\usepackage{amsmath, amsthm} 
\usepackage{multirow} 
\usepackage{amsfonts}
\usepackage{amssymb,graphics,psfrag}
\usepackage{array,epsfig,multirow,stmaryrd,graphicx}
\usepackage{comment}
\usepackage{lscape}
\usepackage{rotating}
\usepackage[all]{xy}
\def\hybrid{\topmargin -20pt    \oddsidemargin 0pt
        \headheight 0pt \headsep 0pt
        \textwidth 6.25in       
        \textheight 9.5in       
        \marginparwidth .875in
        \parskip 5pt plus 1pt   \jot = 1.5ex}
\hybrid
\numberwithin{equation}{section}
\numberwithin{table}{section}

\newcommand{\beq}{\begin{equation}}
\newcommand{\eeq}{\end{equation}}
\newcommand{\be}{\begin{equation}}
\newcommand{\ee}{\end{equation}}
\newcommand{\bea}{\begin{eqnarray}}
\newcommand{\eea}{\end{eqnarray}}   
\newcommand{\ben}{\begin{eqnarray*}}
\newcommand{\een}{\end{eqnarray*}}                  
\newcommand{\ba}{\begin{aligned}}
\newcommand{\ea}{\end{aligned}}
\newcommand{\bt}{\begin{tabular}}
\newcommand{\et}{\end{tabular}}
\newcommand{\bc}{\begin{center}}
\newcommand{\ec}{\end{center}}

\newcommand{\cO}{\mathcal{O}}

\newcommand{\cL}{\mathcal{L}}

\newcommand{\cF}{\mathcal{F}}
\newcommand{\rF}{\mathrm{F}}

\newcommand{\cM}{\mathcal M}

\newcommand{\bk}{\bar{k}}
\newcommand{\bl}{\bar{l}}

\newcommand{\bj}{\bar{j}}

\newcommand{\p}{\partial}

\newcommand{\cref}{{\bf [check ref]}}

\psfrag{n1}{$\nu_1$}
\psfrag{n2}{$\nu_2$}
\psfrag{n1'}{$\nu_1'$}
\psfrag{n2'}{$\nu_2'$}
\psfrag{n9}{$\nu_9$}
\psfrag{n10'}{$\nu_{10}'$}
\psfrag{t1}{$\tilde{\nu}_1$}
\psfrag{t2}{$\tilde{\nu}_2$}
\psfrag{t9}{$\tilde{\nu}_9$}
\psfrag{t1'}{$\tilde{\nu}_1'$}
\psfrag{t2'}{$\tilde{\nu}_2'$}
\psfrag{t10'}{$\tilde{\nu}_{10}'$}
\newtheorem{lem14.1}{Lemma}
\newtheorem{cor14.2}[lem14.1]{Corollary}

\newdimen\tableauside\tableauside=1.0ex
\newdimen\tableaurule\tableaurule=0.4pt
\newdimen\tableaustep
\def\phantomhrule#1{\hbox{\vbox to0pt{\hrule height\tableaurule width#1\vss}}}
\def\phantomvrule#1{\vbox{\hbox to0pt{\vrule width\tableaurule height#1\hss}}}
\def\sqr{\vbox{%
  \phantomhrule\tableaustep
  \hbox{\phantomvrule\tableaustep\kern\tableaustep\phantomvrule\tableaustep}%
  \hbox{\vbox{\phantomhrule\tableauside}\kern-\tableaurule}}}
\def\squares#1{\hbox{\count0=#1\noindent\loop\sqr
  \advance\count0 by-1 \ifnum\count0>0\repeat}}
\def\tableau#1{\vcenter{\offinterlineskip
  \tableaustep=\tableauside\advance\tableaustep by-\tableaurule
  \kern\normallineskip\hbox
    {\kern\normallineskip\vbox
      {\gettableau#1 0 }%
     \kern\normallineskip\kern\tableaurule}%
  \kern\normallineskip\kern\tableaurule}}
\def\gettableau#1{\ifnum#1=0\let\next=\null\else
\squares{#1}\let\next=\gettableau\fi\next}

\def\blfootnote{\xdef\@thefnmark{}\@footnotetext} 
\long\def\symbolfootnote[#1]#2{\begingroup%
\def\thefootnote{\fnsymbol{footnote}}\footnote[#1]{#2}\endgroup}

\begin{document}

\begin{titlepage}

\hfill\vbox{\hbox{Bonn-TH-08-03}}

\vspace*{ 2cm}

\centerline{\Large \bf  Topological Strings on Grassmannian Calabi-Yau manifolds} 

\medskip

\vspace*{4.0ex}

\centerline{\large \rm
Babak Haghighat$^a$ and Albrecht Klemm$^a$
}

\vspace*{4.0ex}
\begin{center}
{\em $^a$\ \ Physikalisches Institut, Universit\"at Bonn, \\[.1cm]
            D-53115 Bonn, BRD}

\vspace*{1.8ex}

\symbolfootnote[0]{\tt \begin{tabular}{ll}$^a$babak@th.physik.uni-bonn.de,
aklemm@th.physik.uni-bonn.de \end{tabular}}

\vskip 0.5cm
\end{center}

\centerline{\bf Abstract}
We present solutions for the higher genus topological string amplitudes  
on Calabi-Yau-manifolds, which are realized as complete intersections in Grassmannians. We
solve the B-model  by  direct integration of the holomorphic anomaly equations
using a finite basis of modular invariant generators, the gap condition at the
conifold and other local boundary conditions for the amplitudes. Regularity of
the latter  at certain points in the moduli space suggests a CFT description. 
The A-model amplitudes are evaluated using a mirror conjecture for
Grassmannian Calabi-Yau  by Batyrev, Ciocan-Fontanine, Kim and Van Straten. 
The integrality of the BPS states gives strong evidence for the conjecture.     

\medskip

\vskip 1cm

\noindent \today
\end{titlepage}

\tableofcontents

\newpage

\section{Introduction}
Mirror symmetry of Calabi-Yau manifolds has been understood to large extend
for complete intersections or hypersurfaces in toric ambient space. 
However a huge and much less explored class of Calabi-Yau manifolds, 
with distinct low energy spectrum, can be realized in ambient spaces, 
which are defined by other homogeneous spaces like the Grassmannians 
$\mathbb{G}(k,n)=U(n)/(U(k)\times U(n-k))$. The topological 
properties of spaces defined by the complex actions of Lie groups 
are described in \cite{BH}.    
From the point of the 2-d linear $\sigma$-model description 
of the ambient space~\cite{wittenphases} the difference is that the former 
have  $U(1)^r$ gauge symmetries, while the latter have non-abelian $\prod_k
U(N_k)$ gauge symmetries.      
The proof that mathematicians~\cite{Kontsevich:1994na} gave for the fact that the $B$-model 
calculation of the genus zero amplitude counts worldsheet instantons 
on the mirror manifold $W$ relies on localisation w.r.t. the $U(1)^r$ action 
and the construction of mirror pairs by reflexive polyhedra. It has 
not been extended to the non-abelian case, e.g. to Grassmannian Calabi-Yau.  
For higher genus amplitudes such proofs are not in general available even on
normal toric ambient spaces, but there are some results on genus one amplitudes 
\cite{Givental}\cite{zinger}. In this article we explore the physical 
mirror symmetry predictions in situations, where it is mathematical 
very difficult to prove along the lines described
above, namely for the higher genus amplitudes on Grassmannian Calabi-Yau 
spaces. Nevertheless the physical integrality conditions on the BPS 
invariants, defined in~\cite{GVInv}\cite{KKV} give strong consistency checks on 
our A-model mirror symmetry predictions on these manifolds.

For genus zero the first steps in the B-model analysis for Grassmannian 
Calabi-Yau spaces have been done in~\cite{BFKS}. Since the usual 
construction of mirror pairs by reflexive polyhedra does apply only to toric
Calabi-Yau, the strategy of the authors is to consider a conifold transition 
from a Grassmannian Calabi-Yau to a toric Calabi-Yau, apply Batyrevs mirror 
construction there and perform an inverse conifold transition back to 
a Grassmannian Calabi-Yau. This is reviewed in section
\ref{mirrorconstruction}. For technical reasons we chose the new  one 
parameter models, for which the mirror geometry and in particular the 
Picard-Fuchs equations were found in \cite{BFKS}. We apply the methods
developed in~\cite{BCOV}\cite{YY}\cite{HKQ} to the B-model. 
Notably the structure of the holomorphic and an-holomorphic 
modular expressions in the amplitudes analysed in~\cite{YY} allows 
for a very effective recursive integration of the holomorphic 
anomaly equations. This structure can be related to the 
traditional theory of holomorphic and anholomorphic modular 
forms of subgroups of $SL(2,\mathbb{Z})$ in the case of 
local mirror symmetry~\cite{Huang:2006si}. For  the large moduli 
space of the Calabi-Yau this formalism can be extended at least formally  
to the global case \cite{Grimm:2007tm}. The automorphic forms 
should be then associated to abelian varieties. 

The direct integration of the holomorphic anomaly  
has to be supplemented with boundary conditions 
to provide the solutions. We find that the gap condition at the 
generic conifold divisor, where an $S^3$ shrinks,  
found in~\cite{HKQ} is present also in the Grassmannian 
Calabi-Yau spaces and provides most of the information. Other boundary 
information is provided by the regularity at CFT points in the 
moduli and places where lens spaces $S^3/\mathbb{Z}_N$ shrink.
The Picard-Fuchs equations of the one parameter Grassmannian
Calabi-Yau spaces are considerably more involved than the 
ones for hypersurfaces and complete intersections in toric 
Calabi-Yau. While the latter have always three regular 
singular points in a $\mathbb{P}^1$ compactification, the former 
have many regular singular points. One motivation for the 
investigation was to analyze the degeneration of the higher 
genus amplitudes at these partly novel singularities and to 
see whether enough boundary conditions can be found to 
solve the theory completely. 
In all one parameter cases, one has been analyzed also in~\cite{HK}, 
we can use the methods described above to solve the model at least 
to genus $5$ and in many cases higher.

\section{Calabi-Yau complete intersections in Grassmannians} \label{GR}

In this section we introduce the Calabi-Yau intersections in Grassmannian,
calculate their topological data and review the mirror construction 
of \cite{BFKS}.

\subsection{Topological invariants of the manifolds}
\label{topologicalintersections} 
Compact Calabi-Yau manifolds $M$ can be constructed by considering complete
intersections in K\"ahler ambient spaces with positive Chern class. 
The first Chern class of the complete intersections is controlled by 
the adjunction formula and we can chose appropriate degrees of the  
complete intersection constraints so that $c_1(TM)=0$. We will 
calculate the topological data of $M$ by basic algebraic geometry.
All necessary tools are reviewed in \cite{GH,BH}.

We restrict to complete intersections in smooth Grassmannians. 
In this way one finds 5 complete intersections $M$ with 
$h^{1,1}=1$.  The ambient 
space will be denoted as $\mathbb{G}(k,n)=(U(k)\times U(n-k))$, 
where $U(n)$ are the unitary groups. For the complete 
intersection we use the notation  
\begin{equation} 
\left(\mathbb{G}(k,n)\right|\!\!\left| d_1,\dots,
  d_l \right)^{h^{1,1}}_\chi \ . 
\end{equation} 
Here the degrees $d_i$ of the Calabi-Yau intersection are given w.r.t. to 
the principal canonical bundle $Q$ of the Grassmannian, see below. 
In addition we give the Euler number $\chi$  as subscript and the Picard number 
$h^{1,1}$ as superscript. Of course, $h_{3,0}=1$, $h_{k,0}=0$ for $k=1,2$  
and $h^{2,1}=-\frac{\chi}{2}+h^{1,1}$. Together with Poincar\'e and 
Hodge duality this fixes all Hodge numbers of $M$. All necessary 
topological data, which fix the topological type of $M$, 
are calculated below using Schubert calculus.

Let us first give a closed expression for the Chern classes of Grassmannians
following Borel and Hirzebruch in \cite{BH}. Their method is based on an 
identification of Chern classes with elementary symmetric
polynomials or combinations of them, which we will summarize here. 

Let $S\{x_1,\cdots, x_l\}$ denote the set of elementary symmetric polynomials in the variables $x_1, \cdots, x_l$. Then
the integral homology $H^*(\mathbb{G}(k,n),\mathbb{Z})$ of the Grassmannian can be identified with the quotient

\begin{equation}
  S\{x_1,\cdots,x_{n-k}\} \otimes S\{x_{n-k+1},\cdots, x_{n}\} / I,
\end{equation}
where $I$ is the ideal generated by the symmetric power series in $x_1, \cdots, x_n$ without constant term. Now, in this 
representation, the closed formula for the total Chern class reads

\begin{equation} \label{cGr}
  c(\mathbb{G}(k,n)) = \prod_{i=1}^{n-k} (1-x_i)^n \prod_{1\leq i \leq j \leq n-k} (1-(x_i-x_j)^2)^{-1}.
\end{equation}

Practically, in order to calculate the Chern classes, substitute each $x_l$ by $h x_l$ and make a series expansion in
$h$. Then, the $i$'s Chern class is given by the coefficient of $h^i$ which can be expressed in terms of elementary 
symmetric polynomials $\sigma_r$, $r\leq i$ in $x_1, \cdots, x_{n-k}$. For
example, we have

\begin{eqnarray}
  c_1 (\mathbb{G}(k,n)) & = & - n \sigma_1, \\ \nonumber
  c_2 (\mathbb{G}(k,n)) & = & \left(\binom{n}{2} + n-k -1 \right) \sigma_1^2 + k \sigma_2.
\end{eqnarray}
The formula for the first Chern class shows that $-\sigma_1$ is a positive generator of $H^2(\mathbb{G}(k,n),\mathbb{Z})$. 
Next, note that $\sigma_r$ is (up to a possible sign) the $r$-th Chern class of the canonical principal $U(n-k)$-bundle $Q$ 
over $\mathbb{G}(k,n)$ and as such represents the class of a hyperplane section. We have $\sigma_1 = -c_1(Q)$, $\sigma_2 =c_2(Q)$,
$\sigma_3=-c_3(Q)$, $\ldots$.

Finally, we are ready to write down the total Chern class of Calabi-Yau complete intersections
$\left(\mathbb{G}(k,n)\right|\!\!\left| d_1,\dots,d_l \right)^{h^{1,1}}_\chi$, $l = k(n-k)-3$, $d_1 + \cdots +d_l = n$ :

\begin{equation}
  c(\left(\mathbb{G}(k,n)\right|\!\!\left| d_1,\dots,d_l \right)^{h^{1,1}}_\chi) = \frac{c(\mathbb{G}(k,n))}{(1+d_1 c_1(Q))\cdots(1+d_l c_1(Q))}.
\end{equation}

Denoting by $H$ the hyperplane $\sigma_1$ , the topological invariants $\chi(M)$, $c_2(M) \cdot H$, $H^3$ can be expressed
through intersection numbers of the Grassmannian $\mathbb{G}(k,n)$. As an example, we review the calculation of the Euler number. The Gauss-Bonnet 
formula gives $\int_M c_3(M) = \chi$. Now, using the adjunction formula, this integral
can be expressed through an integral over the whole Grassmannian

\begin{equation}
  \chi(M) = \int_M c_3(M) = \int_{\mathbb{G}(k,n)} c_3(M) \prod_{i=1}^l d_i H = \int_{\mathbb{G}(k,n)} c_3(M) \prod_{i=1}^l d_i c_1(Q).
\end{equation}

Similarly, the other topological invariants are given by

\begin{equation}
  c_2(M) \cdot H = \int_{\mathbb{G}(k,n)} c_2(M) c_1(Q) \prod_{i=1}^l d_i c_1(Q),
\end{equation}

\begin{equation}
  H^3 = \int_{\mathbb{G}(k,n)} c_1(M)^3 \prod_{i=1}^l d_i c_1(Q).
\end{equation}

As all Chern classes of $M$ are expressed through Chern classes of $Q$, which are Poincare dual to the Schubert cycles of the 
Grassmannian, all invariants can at the end be expressed through intersection numbers of Schubert cycles. These numbers can 
then be calculated utilizing the Schubert calculus and Pieri's formula. Denoting by $\sigma_a$ the special Schubert cycle 
given by the indices $a=(a,0,\cdots,0)$ and by $\sigma_{\underline{b}}$ a general Schubert cycle with indices 
$\underline{b} = (b_1, \cdots, b_k)$, Pieri's formula reads
\\ 
\begin{equation}
  \sigma_a \cdot \sigma_{\underline{b}} = \sum_{\substack{b_i \leq c_i \leq b_{i-1}\\\sum c_i = a + \sum b_i}} \sigma_{\underline{c}}.
\end{equation}

Note that in the above formula the index $c_1$ must always be greater or equal to $b_1$. For further details we refer to \cite{GH}.

We have performed the above steps and list the result for our Calabi-Yau complete intersections in the Appendix.

\subsection{Pl\"ucker embedding} 

In order to describe the mirror of the complete intersections in Grassmannians
it is useful to have an embedding of the Grassmannian into the projective
space. The Pl\"ucker map provides such an embedding. It simply sends 
a $k$-plane $\Lambda=\mathbb{C}\{v_1,\cdots,v_k\}\subset \mathbb{C}^n$ to the multivector $v_1 \wedge \cdots \wedge v_k$.

Explicitly, in terms of the basis $\{e_I= e_{i_1}\wedge \cdots \wedge e_{i_k}\}_{\#I=k}$ for $\wedge^k \mathbb{C}^n$, this
map is given by the data

\begin{equation}
  p : \mathbb{G}(k,n) \rightarrow \mathbb{P}(\wedge^k \mathbb{C}^n)=\mathbb{P}^{\binom{n}{k}-1},
\end{equation}

\begin{equation}
  \Lambda \mapsto [\cdots, |\Lambda_I|,\cdots],
\end{equation}
\\
where the $|\Lambda_I|$ are the determinants of all the $k\times k$ minors of $\Lambda_I$ of a matrix representative
of $\Lambda$.

To describe this embedding algebraically we need to find a set of equations which cut out the Grassmannian in 
$\mathbb{P}^{\binom{n}{k}-1}$, i.e. which define conditions on a multivector $\Lambda \in \wedge^k V$ to be of
the form

\begin{equation}
  \Lambda = v_1 \wedge \cdots \wedge v_k.
\end{equation}
\\
Some calculations show that this is equivalent to demanding

\begin{equation}
  (i(\Xi)\Lambda)\wedge \Lambda = 0,
\end{equation}
\\
for all $\Xi \in \wedge^{k-1} V^{\*}$. Here, the map $i(\Xi)\Lambda$ is defined by

\begin{equation} \label{Plembd}
  \langle i(\Xi) \Lambda,v\rangle~ = ~\langle \Xi,\Lambda \wedge v\rangle
\end{equation}
\\
for all $v \in V$.

Now, a Calabi-Yau complete intersection is obtained by choosing hypersurfaces of appropriate total degree in
$\mathbb{P}^{\binom{n}{k}-1}$, such that their intersection with $\mathbb{G}(n,k)$ is a nonsingular Calabi-Yau space.

\subsection{Mirror Construction}
\label{mirrorconstruction} 
A mirror construction for the above type of Calabi-Yau spaces was given in \cite{BFKS}. Here, we will only sketch 
the method introduced there which is based on conifold transitions. 

Let $M$ be a Grassmannian Calabi-Yau described by the Grassmannian $\mathbb{G}(k,n)$ and hyperplanes $H_i$. As was shown
by Sturmfels \cite{St} a flat deformation of $\mathbb{G}(k,n)$ in its Pluecker embedding leads to a Gorenstein toric Fano variety 
$P(k,n) \subset \mathbb{P}^{\binom{n}{k}-1}$. Now, denote by $M_0$ the intersection of $P(k,n)$ with generic
hypersurfaces $H_i$. This manifold has a locus of conifold singularities which come from the singularities of 
$P(k,n)$. Resolving these by restriction of a small toric resolution of singularities in $P(k,n)$ one obtains a second
Calabi-Yau $M^{*}$. $M^{*}$ is a complete intersection in a toric manifold and as such its mirror construction is known.
The remaining task is to find an appropriate specialization of the toric mirror $W^{*}$ for $M^{*}$ to a conifold 
$W_0$ whose small resolution provides the mirror $W$ of $M$. This task was performed in \cite{BFKS} for the
manifolds we will be dealing with in this paper.

The above steps can be summarized in the following graph:

\setlength{\unitlength}{0.75mm}
\begin{picture}(100,50)
  \put(20,18){$M$}
  \put(30,20){\vector(1,0){40}}
  \put(75,18){$M_0$}
  \put(60,25){\textit{conifold transition}}
  \put(85,20){\vector(1,0){40}}
  \put(130,18){$M^{*}$}
  \put(22,10){\vector(0,-1){40}}
  \put(20,-42){$W$}
  \put(70,-40){\vector(-1,0){40}} 
  \put(75,-42){$W_0$}
  \put(60,-50){\textit{conifold transition}}
  \put(125,-40){\vector(-1,0){40}}
  \put(130,-42){$W^{*}$}
  \put(133,10){\vector(0,-1){40}}
  \put(75,-10){\Large $\circlearrowright$}
\end{picture}

\vspace{50mm}

\section{The BCOV anomaly equation}

In this section, the general procedure for solving the BCOV anomaly equation is reviewed. The connection of the solutions to 
Gromov-Witten potentials is established which will allow us to extract Gopakumar-Vafa invariants from a series expansion of these 
potentials.

\subsection{Special geometry and the topological string}

Here we review how the deformation space of the topological B-model carries the structure of a special K\"ahler manifold
which can be identified with the special K\"ahler geometry of local Calabi-Yau moduli spaces.
As is discussed in \cite{BCOV}, infinitesimal deformations of the topological B-model are parametrized by the chiral fields of 
charge $(q,\bar{q})=(1,1)$. These are the marginal fields which are spanned by a basis $\phi_i$ for $i=1, \cdots, n$. In fact, the 
deformations span a complex manifold $\cM$ of dimension $n$. We are interested
in the ring spanned by $(\phi_0,\phi_i, \phi^i, \phi^0)$, where $\phi_0$ is the identity operator of charge $(q,\bar{q})=(0,0)$, and
$\phi^i$ are the charge $(2,2)$ fields and finally $\phi^0$ is the top element in the chiral ring of charge $(3,3)$. These fields satisfy
the following identities with respect to the topological metric

\begin{equation}
  \eta(\phi_i,\phi^j) = \langle \phi_i \phi^j\rangle_0 = \delta^j_i,
\end{equation}

\begin{equation}
  \eta(\phi_0,\phi^0) = \langle \phi_0 \phi^0 \rangle_0 = 1.
\end{equation}

Here, $\langle \cdot \rangle_0$ denotes the topological correlation function on the sphere. 
The ring structure is encoded in the so called Yukawa coupling, which is the three-point function on the sphere

\begin{equation}
  C_{ijk} = \langle \phi_i \phi_j \phi_k\rangle_0.
\end{equation}

Using the operator state correspondence one can define 

\begin{equation}
  |i\rangle = \phi_i |0\rangle,
\end{equation}
and the topological metric becomes

\begin{equation}
  \eta(\phi_i, \phi^j) = \langle i | j\rangle = \delta^j_i.
\end{equation}

Finally, one can define a hermitian metric using the worldsheet CPT operator $\Theta$,

\begin{equation}
  g_{i \bar{j}} = \langle \Theta j | i\rangle.
\end{equation}

Now, moving around in the moduli space $\cM$, the space of states generated by the chiral fields forms a holomorphic vector bundle $V \rightarrow \cM$.
It can be shown that its charge $(0,0)$ subspace forms a holomorphic line bundle $\cL$ over $\cM$ and that the
charge $(1,1)$ subbundle corresponds to the line bundle $\cL \times T \cM$. The charge $(2,2)$ and $(3,3)$ subbundles respectively turn out to be duals
of $\cL \times T \cM$ and $\cL$. These bundles are described through their covariant derivatives which will be given in the following.
On $\cM$ one can define a metric, called Zamolodchikov metric, 

\begin{equation} \label{spK1}
  G_{i \bar{j}} = \frac{g_{i \bar{j}}}{g_{0 \bar{0}}},
\end{equation}
which is K\"ahler with K\"ahler potential $K = - \log g_{0 \bar{0}}$. The connections on $\cL$ and $T\cM$ are now given by $\partial_i K$ 
and the metric connection $\Gamma^i_{jk}$ for $G_{i \bar{j}}$. The covariant derivative of a section $\xi \in \Gamma(\cL^n \times T \cM^m)$ 
is then given by 

\begin{equation}
  D_i \xi^{j_1 \cdots j_m} = \partial_i \xi^{j_1 \cdots j_m} + \Gamma^{j_1}_{ik} \xi^{k~j_2 \cdots j_m} 
                              + \cdots \Gamma^{j_m}_{ik} \xi^{j_1 \cdots j_{m-1} k} + n \partial_i K \xi^{j_1 \cdots j_m}.
\end{equation} 

In this picture, the Yukawa coupling is a symmetric rank 3 tensor with values in $\cL^{2}$, which furthermore obeys the constrains

\begin{equation} \label{spK2}
  \partial_{\bar{l}} C_{ijk} = 0, ~~~~  D_i C_{jkl} = D_j C_{ikl}.
\end{equation}

Finally, for the curvature of the Zamolodchikov metric one obtains the relation

\begin{equation} \label{spK3}
  (R_{i\bar{j}})^k_l = [D_i,D_{\bar{j}}]^k_l = C_{ilm} C_{\bar{i}\bar{m}\bar{k}} e^{2K} G^{m\bar{m}} G^{k \bar{k}} 
                       - \delta^k_l G_{i\bar{j}} - \delta^k_i G_{l\bar{j}}.
\end{equation}

The equations \ref{spK1}, \ref{spK2} and \ref{spK3} define the so called special K\"ahler geometry.

A Calabi-Yau threefold can be defined as K\"ahler manifold, which has a
no-where vanishing $(3,0)$ form $\Omega({\underline z})$, depending on the
complex structure deformations ${\underline z}$. We denote the mirror of $M$ on
which we evaluate the periods by $W$. One has simple formulas for the
K\"ahler  potential $K$  and the Yukawa couplings $C_{ijk}$ in terms of 
integrals over $W$. In particular  
\begin{equation}
e^{-K}=i \int_W \Omega \wedge \bar \Omega\ =: (\Omega,\bar \Omega) \  
\end{equation} 
and         
\begin{equation}
C_{ijk} =\int_W \Omega \wedge \partial_{z_i}\partial_{z_j}\partial_{z_j}  \Omega\ . 
\end{equation}
One can reduce these integrals to period integrals and ultimately to certain
solutions of the Picard-Fuchs equation as follows. First one chooses an integral 
symplectic basis $\{A^k,B_k\}$, $k=1,\ldots,h_{2,1}(W)+1$ of three cycles 
in $H_3(W,\mathbb{Z})$, i.e. $A^k\cap B_l=\delta^k_l$ such that all other
intersections are zero, see  \cite{Ca}. Then one chooses a dual 
basis $\{\alpha_l,\beta^k\}$, $k=1,\ldots,h_{2,1}(W)+1$ of three forms in
$H^3(W,\mathbb{Z})$. 
It fulfills $\int_{A^k}\alpha_l=\delta^k_l$, $\int_{B_k}\beta^l=\delta^l_k$,
while all other pairings are zero. One has $(\alpha_l,\beta^k)=i\delta^k_l$,
while  again all other pairings are zero. Now we can expand 
\begin{equation} 
\Omega(z)=X^k(z)\alpha_k-F_l(z)\beta^l
\end{equation}  
in terms of the periods $X^k(z)=\int_{A^k}\Omega(z)$ as well  $F_k(z)=\int_{B_k}\Omega(z)$. 

To recover the period integrals over the basis $\{A^k,B_k\}$
from the solutions  of the Picard-Fuchs equations we use special geometry and 
the typical degeneration of the periods at the point of maximal unipotent monodromy.  
First we note that the $X^k$  serve as homogenous coordinates for the space 
of complex structures. As a consequence of Griffith transversality 
$F^{(0)}(X^k):=\frac{1}{2} X^k F_k(X^k)$ is homogenous of degree $2$ in $X^k$ 
and $F_k=\partial_{X^k} F^{(0)}$. $F^{(0)}$ is called the prepotential. At the
point of maximal unipotent monodromy we have
\begin{equation}
\vec \Pi  =
\left(
\begin{array}{c}
\int_{B_1} \Omega\\
\int_{B_2} \Omega\\
\int_{A^1} \Omega\\
\int_{A^2} \Omega
\end{array}
\right)=
\left(
\begin{array}{c}
F_0\\
F_1\\
X_0\\
X_1\\
\end{array}
\right)=
\omega_0\left(
\begin{array}{c}
2{\cal F}^{(0)} -t \partial_t {\cal F}^{(0)}\\
\partial_t {\cal F}^{(0)} \\
1\\
t\\
\end{array}
\right)=\left(
\begin{array}{c}
\omega_3+c\, \omega_1+e\, \omega_0 \\
-\omega_2- a\, \omega_1+c \, \omega_0 \\
\omega_0\\
\omega_1 \\
\end{array}
\right)\ ,
\label{periods}
\end{equation}
where $\omega_0$ is the unique power series solution and $\omega_k$ are
solutions, which behave like $\omega_0(z) \log(z)^k$ at infinity. The 
Frobenius method gives a canonical basis of these
solutions. $t=\frac{\omega_1}{\omega_0}$ is the mirror map 
and in terms of the latter the prepotential looks as follows
\begin{equation}
{\cal F}^{(0)}=-{\kappa\over 3!} t^3-{a\over 2}  t^2+ c t+\frac{e}{2}+
f_{inst}(q) \ ,
\label{prepotential}
\end{equation} 
where $\kappa=H^3$, $c=\frac{1}{24}\int_M c_2 \wedge H$,
$e=\frac{\zeta(3)\chi(M)}{(2 \pi i)^3}$ and 
$a=\frac{1}{2}\int_M i_* c_1(H)\wedge H$. 
All these numbers are calculated on $M$ using the formalism  in 
section \ref{topologicalintersections} and they fix the integral 
symplectic basis on $W$ completely.

\subsection{General solutions of the BCOV anomaly equation} 

The special geometry relations $\bar{\partial}_i C_{jkl}=0$ and $D_i C_{jkl}=D_j C_{ikl}$ allow us to
integrate the Yukawa coupling and its complex conjugate and express them through potential functions

\begin{equation} \label{F0d}
  C_{jkl} = D_j D_k D_l \mathcal{F}^{(0)},
            ~~~C_{\bj\bk\bl}=D_{\bj} D_{\bk} D_{\bl}\mathcal{\bar{F}}^{(0)}.
\end{equation}

Here, $\mathcal{F}_0^{(0)}$ is a $C^{\infty}$ section of $\mathcal{L}^2$ as $C_{jkl}$ is such a section.
Analogously, $\bar{\mathcal{F}}^{(0)}$ is a $C^{\infty}$ section of $\bar{\mathcal{L}}^2$. In the one moduli cases
we are considering here equation \ref{F0d} turns into

\begin{equation}
  C_{zzz}=D_z D_z D_z \mathcal{F}^{(0)}(z,\bar{z}),~~~C_{\bar{z}\bar{z}\bar{z}}=D_{\bar{z}} D_{\bar{z}} D_{\bar{z}} \bar{\mathcal{F}}^{(0)}(z,\bar{z}).
\end{equation}

The genus one free energy suffers from a holomorphic anomaly first calculated in \cite{BCOV1},

\begin{equation} \label{g1ae}
  \bar{\p}_{\bar{k}} \p_m \cF^{(1)}=\frac{1}{2} \bar{C}_{\bar{k}}^{ij} C_{mij} 
                            - (\frac{\chi}{24}-1) G_{\bar{k} m}.
\end{equation}

This equation can be integrated straightforward and one obtains 

\begin{equation} \label{g1fe}
  \cF^{(1)}(z) = \log(\det(G^{-1})^{\frac{1}{2}} e^{\frac{K}{2} (3 + h^{2,1} - \frac{1}{12} \chi)} |f_1|^2),
\end{equation}
where the holomorphic ambiguity is of the form

\begin{equation} \label{g1ha}
  f_1(z) = \prod_i \Delta_i^{r_i} \prod_{i=1}^{h_{21}} z_i^{c_i}.
\end{equation}

Here the $\Delta_i$ are the components of the discriminant and the constants $r_i$ and $c_i$ are determined from
the boundary behavior. In case of the conifold component of the discriminant $\Delta_{con}$ the constant $r_{con}$ is
universally given by $\frac{1}{12}$ as was first pointed out in \cite{VonC}. The $c_i$ are fixed by requiring the boundary 
condition

\begin{equation} \label{lvl}
  \lim_{z_i\rightarrow 0} \cF^{(1)} = -\frac{1}{24} t_i \int_M c_2 \cdot H.
\end{equation}

The higher genus generalization of the holomorphic anomaly is given through a recursion relation, the BCOV holomorphic 
anomaly equation (\cite{BCOV}),

\begin{equation} \label{hgae}
  \bar{\p}_{\bar{k}} \cF^{(g)} = \frac{1}{2} \bar{C}^{ij}_{\bar{k}} \left(D_i D_j \cF^{(g-1)} 
                   + \sum^{g-1}_{r=1} D_j \cF^{(g-1)} D_i \cF^{(r)}\right),
\end{equation}

where the $\cF^{(g)}$ are $C^{\infty}$ sections of $\cL^{2-2g}$.

The idea presented in \cite{BCOV} to solve this equation is to rewrite the right hand side as a derivative with respect to
$\bar{\p}_{\bar{k}}$

\begin{eqnarray}
  \bar{\p}_{\bar{k}} \cF^{(g)} &=& \bar{\p}_{\bar{k}} \left(\frac{1}{2} S^{ij} \left(D_i D_j \cF^{(g-1)}
                                  + \sum_{r=1}^{g-1} D_i \cF^{(r)} D_j \cF^{(g-1)}\right)\right) \nonumber \\
                              &~& -\frac{1}{2}S^{ij} \bar{\p}_{\bar{k}}\left(D_i D_j \cF^{(g-1)} 
                                  +\sum_{r=1}^{g-1} D_i \cF^{(r)} D_j \cF^{(g-r)}\right),
\end{eqnarray}

where $S^{ij}$ is implicitly defined through

\begin{equation}
  \bar{C}^{ij}_{\bar{k}}=\bar{\p}_{\bar{k}}S^{ij}.
\end{equation}

Using the commutator 

\begin{equation}
  R^l_{i\bar{k}j}=[D_i,\bar{\p}_{\bar{k}}]^l_j=G_{i\bar{k}} \delta^l_j + G_{j\bar{k}} \delta^l_i
                  -C^{(0)}_{ijm} \bar{C}^{ml}_{\bar{k}}
\end{equation}

allows one to rewrite the second term in such a way that the $\bar{\p}_{\bar{k}}$ derivative acts directly on the 
$\cF^{(g)}$. Then the holomorphic anomaly equations for $g'<g$ can be used iteratively to generate an equation of the form

\begin{equation}
  \bar{\p}_{\bar{k}} \cF^{(g)}=\bar{\p}_{\bar{k}}\Gamma^{(g)}(S^{ij},S^i,S,C^{(<g)}_{i_1,\cdots,i_n}),
\end{equation}

where $S^i$, $S$ and $C^{(<g)}_{i_1,\cdots,i_n}$ are defined through

\begin{equation}
  \bar{C}_{\bar{j}\bar{k}\bar{l}}=e^{-2K} \bar{D}_{\bar{i}}\bar{D}_{\bar{j}}\bar{D}_{\bar{k}}S,~~
  S_{\bar{i}}= \bar{\p}_{\bar{i}} S,~~ S^j = G^{j \bar{k}} S_{\bar{k}},
  ~~C^{(g)}_{i_1,\cdots,i_n}=D_{i_1}\cdots D_{i_n}\cF^{(g)}.
\end{equation}

A solution is given by

\begin{equation} \label{gengfe}
  \cF^{(g)}=\Gamma^{(g)}(S^{ij},S^i,S,C^{(<g)}_{i_1,\cdots,i_n})+f^{(g)}.
\end{equation}

where $f^{(g)}$ is the holomorphic ambiguity, which is not fixed by the recursive procedure. The method we will use
to fix this ambiguity genus by genus is to go to boundary points of moduli space and use physical interpretation at those
points to reconstruct the ambiguity globally. However, it is important to note that the boundary information is not
restrictive enough to carry out the procedure up to genus infinity.

\subsection{Topological limit and Gromov-Witten potentials}

The topological limit of the free energy was introduced in \cite{BCOV1}. In order to define it we first have to introduce 
the normalized solutions of the Picard-Fuchs equation around the large volume point in moduli space. As we are dealing
with one parameter models, let these be given by $\omega_0(z)$ and $\omega_1(z)$, which determines the mirror map to be 
$t=t(z)=\frac{\omega_1(z)}{\omega_0(z)}$. With these notations we can now introduce the topological limit to be defined 
by the following replacements,

\begin{equation}
  G_{z\bar{z}}\rightarrow \frac{dt}{dz}\frac{d\bar{t}}{d\bar{z}},~~K_z \rightarrow -\p_z \log \omega_0(z),
  ~~\cF^{(g)}(z,\bar{z})\rightarrow F^{(g)}(z),
\end{equation}
in the solution \ref{gengfe}, giving

\begin{equation}
  F^{(g)}(z) = \Gamma(S^{zz}(z),S^z(z),S,C_r^{(<g)})+f_g(z).
\end{equation}

This determines the $F^{(g)}$ to be holomorphic prepotentials and sections of $\cL^{(2-2g)}$. The Gromov-Witten
potential is given through this holomorphic prepotential by

\begin{eqnarray} \label{gwp}
  \rF_g(t) &=& (\omega_0(z))^{2g-2} F^{(g)}(z) \nonumber \\
           &=& (\omega_0(z))^{2g-2} \Gamma(S^{zz},S^z,S,C^{(<g)}_r(z)) + (\omega_0(z))^{2g-2} f_g(z).
\end{eqnarray}

This function is the generating function of the Gromov-Witten invariants $N_g(d)$ and its expansion in terms of these is
given by

\begin{equation}
  \rF_g(t) = \frac{\chi}{2} (-1)^g \frac{|B_{2g} B_{2g-2}|}{2g (2g-2) (2g-2)!} + \sum_{d>0} N_g(d) q^d,~(q=e^{2 \pi i t}),
\end{equation}
where $\chi$ is the Euler number of the Calabi-Yau manifold and $B_g$ is the $g$th Bernoulli number. 

For applications to the enumerative problem of counting holomorphic curves and/or the extraction of the physical
content in terms of BPS states it is reasonable to switch to the effective action point of view. From this point of
view the series $F(\lambda,t) = \sum_{g=1}^{\infty} \lambda^{2g-2} \rF^{(g)}(t)$ computes the following term in the
effective $N=2$ superpotential:

\begin{equation}
  S^{N=2}_{1-loop} = \int d^4 x R_{+}^2 F(\lambda,t),
\end{equation}
where $R_{+}$ is the self-dual part of the curvature and $\lambda$ is identified with the self-dual part
of the graviphoton field strength $F_{+}$. Alternatively, this term is calculated by a one-loop integral in a constant
graviphoton background, where the particles running in the loop are charged BPS states. The calculation is very similar
to the ordinary Schwinger-loop calculation and the result is 

\begin{equation}
  \sum_{g \geq 0} \lambda^{2g-2} \rF_g(t) = \sum_{g\geq 0} \sum_{k\geq 1,d \geq 0} n_g(d) \frac{1}{k}
                                            (2 \sin \frac{k \lambda}{2})^{2g-2} q^{kd}.
\end{equation}

The $n_g(d)$ are the so-called Gopakumar-Vafa invariants and are integral.

\newpage

\section{The Grassmannian Calabi-Yau $\left(\mathbb{G}(2,5)\right|\!\!\left| 1,1,3 \right)^{1}_{-150}$}

This Calabi-Yau manifold is obtained as a complete intersection of hypersurfaces in the Grassmannian $\mathbb{G}(2,5)$ as described in section \ref{GR}. 
In our special case the Pl\"ucker embedding is an embedding of $\mathbb{G}(2,5)$ into $\mathbb{P}^9$ and equations \ref{Plembd} take the form

\begin{eqnarray}
  z_{23} z_{45} - z_{24} z_{35} + z_{25} z_{34} & = & 0 , \nonumber \\
  z_{13} z_{45} - z_{14} z_{35} + z_{15} z_{34} & = & 0 , \nonumber \\
  z_{12} z_{45} - z_{14} z_{35} + z_{15} z_{34} & = & 0 , \nonumber \\
  z_{12} z_{35} - z_{13} z_{25} + z_{15} z_{23} & = & 0 , \nonumber \\
  z_{12} z_{34} - z_{13} z_{24} + z_{14} z_{23} & = & 0 .
\end{eqnarray}

Now, the Calabi-Yau $\left(\mathbb{G}(2,5)\right|\!\!\left| 1,1,3 \right)^{1}_{-150}$ is defined to be a smooth 3-dimensional Calabi-Yau complete 
intersection of 3 hypersurfaces of degrees $1,1$ and $3$ in $\mathbb{P}^9$ with $\mathbb{G}(2,5)$.
A calculation shows that we have $h^{1,1}=1$, $h^{2,1}=76$ and $\chi(M)=-150$.

\subsection{Picard-Fuchs differential equations and the structure of the moduli space}

The Picard-Fuchs operator is given by:

\begin{eqnarray} \label{PFgr113i25}
  \mathcal{P} & = & -18 z - 360 z^2 + (-147 z - 2106 z^2) \theta + (-444 z - 3969 z^2) \theta^2 \\ \nonumber
              & ~ & + (-594 z - 2916 z^2) \theta^3 + (1-297 z - 729 z^2) \theta^4,
\end{eqnarray}
where $\theta=z \frac{d}{dz}$. As one can read off, the discriminant is given by $\textrm{dis}(z) = 1-297 z - 729 z^2$.
The Yukawa coupling can be extracted from the Picard-Fuchs operator and its normalization
is determined by the intersection number $H^3$ given in section 2. This procedure is explained in \cite{Ca} and the result for
our particular example is

\begin{equation}
  C_{zzz} = \frac{15}{z^3 (1-11 \cdot 3^3 z - 3^9 z^2)}.
\end{equation}

We expect the solutions to develop logarithmic singularities around the points 
$\textrm{dis}(\alpha_i) = 0,~i\in \{1,2\}$, which indeed occur as can be seen from the 
index structure at these points: 

\begin{equation} \label{pfidx}
  (\rho_1, \rho_2, \rho_3, \rho_4) = (0,1,1,2).
\end{equation}

These points are known as the conifold-points of the moduli space. As is known through the work of Strominger \cite{Str} at 
these points certain non-perturbative type II RR-states become massless and integrating them out leads to singularities 
in the Wilsonian effective action. Such a singularity occurs also in the free energies of the topological string, as 
was first observed in \cite{VonC}, as these free energies calculate couplings of the four dimensional effective field theory. 
While calculating genus g topological string amplitudes we will make extensive use of the knowledge that such massless 
states exist to put restrictive bounds on the holomorphic ambiguity. 

Another special point in our particular moduli space is the point at infinity. Here the Picard-Fuchs-operator develops 
the following indices: 
$(\rho_1, \rho_2, \rho_3, \rho_4) = (\frac{1}{3}, \frac{2}{3}, \frac{4}{3}, \frac{5}{3})$. The $\mathbb{Z}_3$-symmetry 
at this point suggests that it is the enhanced symmetry point of a particular Landau-Ginzburg orbifold model. 
Putting regularity conditions on topological string free energies at this point gives us another bound on the holomorphic 
ambiguity and the resulting Gopakumar-Vafa invariants will give us a consistency check whether 
our regularity assumption was justified.

Finally, the structure of the singularities can be summarized in the following table

\begin{center}

\begin{tabular}{c|cccc} \label{pfidx}
  z & 0 & $\alpha_1$ & $\alpha_2$ & $\infty$\\
  \hline \\
  $\rho_1$ & 0 & 0 & 0 & 1/3 \\
  $\rho_2$ & 0 & 1 & 1 & 2/3 \\
  $\rho_3$ & 0 & 1 & 1 & 4/3 \\
  $\rho_4$ & 0 & 2 & 2 & 5/3 \\
\end{tabular}

\end{center}

\subsection{$g=0$ and $g=1$ Gopakumar-Vafa invariants}

In this section we summarize the calculations of the genus zero and one Gopakumar-Vafa invariants for the Grassmannian. 
We will solve the Picard-Fuchs equation around the point $z=0$ and obtain the mirror map at this point. 

The normalized regular solution and the linear-logarithmic solution are 

\begin{equation}
  \left.
  \begin{array}{ccl}
    \omega_0(z) & = & 1 + 18 z + 1710 z^2 + 246960 z^3 + 43347150 z^4 + \cdots\\
    \omega_1(z) & = & \log x_0(z) + 75 z + \frac{16497}{2} z^2 + 1257046 z^3 + \frac{907324065}{4} z^4 + \cdots
  \end{array}
  \right\}
\end{equation}

The complexified K\"ahler modulus is defined through $2 \pi i t=\frac{\omega_1(z)}{\omega_0(z)}$ and the $q$-expansion of the 
$z$-coordinate takes the following form:

\begin{equation}
  z = q - 75 q^2 + 1539 q^3 - 60073 q^4 + \cdots,
\end{equation}
where $q := e^{2 \pi i t}$.

Now, we are able to determine the quantum corrected Yukawa coupling $K_{ttt}(t)$ at $z=0$. It is given by

\begin{equation}
  \left(\frac{1}{\omega_0(z)}\right)^2 C_{zzz} \left(\frac{dz}{dt}\right)^3 
  = 15 + 540 q + 100980 q^2 + 16776045 q^3 + 2873237940 q^4 + \cdots.
\end{equation}

From these Yukawa couplings we can obtain the Gromov-Witten potential 

\begin{equation}
  K_{ttt}(t) = \left(q \frac{d}{dq}\right)^3 F_0(t).
\end{equation}

The genus one invariants are obtained through the BCOV formula for the holomorphic potential which is the
topological limit of \ref{g1fe} 

\begin{equation}
  F^{(1)}(z) = \frac{1}{2} \log \left\{\left(\frac{1}{\omega_0(z)}\right)^{3+h^{1,1}
               -\frac{\chi}{12}} \left(\frac{dz}{dt}\right) dis(z)^{-\frac{1}{6}}
                z^{c-1-\frac{c2\cdot H}{12}}\right\},
\end{equation}
where we determine $c=0$ through the boundary behavior \ref{lvl}. As both zeros of the discriminant describe
conifold points , it appears with factor $-1/12$ in the logarithm. 

Using the mirror map $z = z(q)$ we finally obtain the genus one Gromov-Witten potentials

\begin{equation}
  \mathrm{F}^M_1(t) = F^{(1)}(z(q)).
\end{equation}

\subsection{Higher genus GV-Invariants}

In this section we explain the recursive solution of the BCOV holomorphic anomaly equation
found in \cite{YY} utilizing the polynomial structure of the partition functions. The topological limits at
certain points in the moduli space are calculated giving boundary conditions on the holomorphic
ambiguity.

\subsubsection{Recursive Solution of the BCOV equation}

The general form \ref{gengfe} of the solution to \ref{hgae} is not so useful for higher genus calculations
as the procedure to determine the anholomorphic part grows exponentially with the genus. The situation can be improved
once one notices that the terms appearing in the Feynman graph expansion are not completely independent, as was
first observed in \cite{KKV}. Using these interrelations, in \cite{YY} a recursive procedure for the quintic was developed whose 
complexity grows only polynomially with the genus.

The basic idea is to introduce two sets of generators, given by

\begin{equation} \label{ABgnr}
  A_k = G^{z \bar{z}} \theta^k_z G_{z\bar{z}},~~B_k = e^{K(z,\bar{z})} \theta^k_z e^{-K(z,\bar{z})},
\end{equation}
where $\theta_z = z \frac{d}{dz}$. A short calculation shows

\begin{equation}
  \theta_z A_k = A_{k+1} - A_1 A_k, ~~ \theta_z B_k = B_{k+1} - B_1 B_k.
\end{equation}

Noticing the relation $e^{-K(z,\bar{z})}=(\Omega(z),\bar{\Omega}(z))$, the Picard-Fuchs equation corresponding to the 
Picard-Fuchs operator \ref{PFgr113i25} can be rewritten in terms of the $B_k$

\begin{equation} \label{Btrc}
  B_4 = r_1(z) B_1 + r_2(z) B_2 + r_3(z) B_3 + r_4(z),
\end{equation}
where the $r_k(z)$ are rational functions. 

Furthermore, there exists a similar relation for the $A_k$. As was shown in \cite{YY} $A_2$ is given by 

\begin{equation} \label{A2trc}
  A_2 = -4 B_2 - 2 B_1 (A_1 - B_1 - 1) + \theta_z \textrm{log}(z C_{zzz}) T_{zz} + r(z),
\end{equation}
where $T_{zz}$ is defined through the $S^{zz}$ propagator 

\begin{equation}
  T_{zz} = -(z C_{zzz}) S^{zz},
\end{equation}
and $r(z)$ is a holomorphic function to be specified later. Also the propagators are defined up to holomorphic functions $f$ and $v$ 

\begin{eqnarray*}
  S^{zz} & = & \frac{1}{C_{zzz}} \left( 2 \partial \textrm{log} (e^{K} |f|^2) - (G_{z \bar{z}} v)^{-1} \partial (v G_{z \bar{z}})\right) \nonumber \\
         & = & -\frac{1}{z C_{zzz}} \left(2 B_1 + 2 \frac{\partial f}{f} + A_1 - \frac{\partial v}{v} \right).
\end{eqnarray*}

We will make a choice of $f$ and $v$, such that the invariant combinations $e^K |f|^2$ and $G_{z \bar{z}} |v|^2$ remain finite around $z=0$. 
The calculation is most conveniently performed by taking the topological limit and we obtain $v=z$ and $f=1$. Therefore, $T_{zz}$
takes the form

\begin{equation}
  T_{zz} = 2 B_1 + A_1 + 1.
\end{equation}

The rational function $r(z)$ is obtained by taking the topological limit of both sides of equation \ref{A2trc} and making
the Ansatz

\begin{equation}
  r(z) = c_0 + c_1 \frac{1}{dis(z)} + c_2 \frac{z}{dis(z)}.
\end{equation}

The coefficients $c_i$ are extracted by comparing both sides and we obtain

\begin{equation}
  r(z) = -\frac{4}{9} + \frac{13}{9 (1 - 297 z - 729 z^2)} - \frac{282 z}{(1 - 297 z - 729 z^2)}.
\end{equation}

The two equations \ref{A2trc} and \ref{Btrc} show that the $\theta_z$-derivative acts within the ring 
generated by $A_1$,$B_1$,$B_2$ and $B_3$. More precisely, we have the property

\begin{equation}
    \theta_z : \mathbb{C}(z)[A_1,B_1,B_2,B_3] \rightarrow \mathbb{C}(z)[A_1,B_1,B_2,B_3].
\end{equation}

Similarly, the action of the $\p_{\bar{z}}$ derivative just adds two more generators to the above polynomial ring, namely
$\p_{\bar{z}}B_1$ and $\p_{\bar{z}}A_1$. This is because, as was shown in \cite{YY} as well as in \cite{HK}, one has the 
following identities

\begin{equation}
  \p_{\bar{z}}B_2 = (1+A_1 + 2 B_1) \p_{\bar{z}},
\end{equation}

\begin{equation}
  \p_{\bar{z}} B_3 = (A_2 + 3 B_1 + 3 B_2 + 3 A_1 B_1 + 1) \p_{\bar{z}} B_1.
\end{equation}

The next step will be to show that rewriting the holomorphic anomaly equations allows us to rewrite the solutions in terms
of polynomials in $A_1$, $B_1$, $B_2$ and $B_3$. In order to proceed we first introduce the quantities $P^{(g)}_n$ defined
through

\begin{equation}
  P^{(g)}_n = (z^3 C_{zzz})^{g-1} z^n D^n_z \mathcal{F}^{(g)}~~~(n=0,1,2,\ldots).
\end{equation}

Under the assumption that $\partial_{\bar{z}}A_1$, $\partial_{\bar{z}}B_1$ are independent the BCOV equation

\begin{equation}
  \partial_{\bar z} P^{(g)} = \frac{1}{2} \partial_{\bar z} (z C_{zzz} S^{zz}) \left\{P^{(g-1)}_2 + \sum^{(g-1)}_{r=1} P^{g-1}_1 P^{(r)}_1\right\}
\end{equation}
can be translated into 
   
\begin{eqnarray*}
  0 & = & 2 \frac{\partial P^{(g)}}{\partial A_1} - \Big(\frac{\partial P^{(g)}}{\partial B_1} + \frac{\partial_{\bar z} B_2}{\partial_{\bar z} B_1}
  \frac{\partial P^{(g)}}{\partial B_2} +  \frac{\partial_{\bar{z}} B_3}{\partial_{\bar{z}} B_1}\frac{\partial P^{(g)}}{\partial B_3}\Big), \nonumber \\
  \frac{\partial P^{(g)}}{\partial A_1} & = & -\frac{1}{2} \left\{P_2^{g-1} + \sum^{g-1}_{r=1} P_1^{(g-r)} P_1^{(r)}\right\}.
\end{eqnarray*}

This shows the polynomiality of the solutions. Performing the following variable change

\begin{eqnarray*}
  u  & = & B_1,~~~~v_1=1+A_1 + 2 B_1,~~~~v_2=-B_1 - A_1 B_1 - 2 B_1^2 + B_2,~~~~\\
  v_3 & = &-B_1 - 2 A_1 B_1 - 5 B_1^2 - A_1 B_1^2 - 2 B_1^3 + B_1 B_2 + B_3 \\
     & ~ & - B_1 (r(z) + T_{zz} \theta_z \textrm{log}(z C_zzz)),
\end{eqnarray*}
one can furthermore obtain $\frac{\partial}{\partial u} P^{(g)}=0$ which reduces the number of independent variables to 
three. Notice that the above equations are generic for all kinds of one parameter models, once $r(z)$ is extracted from
the truncation relation \ref{A2trc}. The holomorphic anomaly equation can now be solved recursively with the initial 
data $P^{(0)}_3=1$ and $P^{(1)}_1$, given by

\begin{equation}
  P^{(1)}_1 = \frac{1}{2} \left\{-A_1 - (2 + h^{11}-\frac{\chi}{12})B_1 - 1 - \frac{c_2 \cdot H}{12}
              - \frac{\theta_z(dis(z))}{6~dis(z)}\right\}.
\end{equation}

A nice way to perform the integration is given in \cite{HK}.

However, the integration of the holomorphic anomaly still leaves us with the holomorphic ambiguity. The relation between
the genus g free energy $\mathcal{F}^{(g)}$, the holomorphic ambiguity $f_g(z)$ and the polynomials $P^{(g)}$ is given by 
the following equation

\begin{equation} \label{FPf}
  \mathcal{F}^{(g)} = (z^3 C_zzz)^{(1-g)} P^{(g)} + f_g(z).
\end{equation}

The Gromov-Witten potentials are once again obtained through equation \ref{gwp}, where one has to make the substitutions

\begin{equation}
  A_1 \rightarrow \Big(\frac{dz}{dt}\Big) \theta_z \Big(\frac{dt}{dz}\Big), 
  ~~B_k \rightarrow \frac{1}{\omega_0(z)} \theta_z^k \omega_0(z),
\end{equation}
in the polynomial solutions $\cF^{(g)}=\cF^{(g)}(A_1(z,\bar{z}),B_k(z,\bar{z}),z)$.

\subsection{Holomorphic ambiguity and boundary conditions}

Requiring regularity of $\rF_g(t)$ at $z=0$ and $z=\infty$, we parameterize the holomorphic ambiguity through the Ansatz

\begin{equation}
  f_g(z) = a_0 + a_1 z + \cdots + a_{2 g - 2} z^{2g-2} 
           + \frac{c_0 + c_1 z + \cdots + c_{4g-5} z^{4g-5}}{dis(z)^{2g-2}}.
\end{equation}

From this we see that the total number of unknown parameters is $6 (g-1) + 1$ and grows linearly in $g$.

One of the main conceptual problems of topological string theory on compact Calabi-Yau is the determination of the 
holomorphic ambiguity. Boundary conditions may be given through the effective 4d action, but also, in some cases, 
geometrical considerations can be of use. For example, we can utilize the first few $N_g(d)$ in the expansion of the 
Gromov-Witten potential once they are known through geometrical calculations. Usually, one puts the lower degree 
Gopakumar-Vafa invariants $n_g(d)$ to zero as they count the number of genus $g$ holomorphic curves in the Calabi-Yau.
Once one knows that the $n_g(d)$ are vanishing up a certain degree for a specific genus $g$, then one knows that 
they must be zero at least up to the same degree for genus $g+1$. This knowledge one can impose as boundary condition
for the Gromov-Witten potentials.
As boundary conditions from physical considerations are far more restrictive for higher genus calculations we will 
concentrate on these in this paper. In order to fix the ambiguity we evaluate the Gromov-Witten potentials at special 
points on the moduli space, where the physics is sufficiently well understood. 

\subsubsection{Expansion around the conifold points}

Our model admits two conifold points and as was first observed in \cite{HKQ} each of them provides us with a gap-like 
structure in the higher genus topological string amplitudes which in turn impose $2g-2$ conditions on the holomorphic 
ambiguity. 

In order to make use of the gap condition we have to compute the topological limit around each conifold singularity. We
denote the conifold singularity by $c$, i.e. in our case $c$ stands for either $\alpha_1=1/54 (-11-5 \sqrt{5})$ or 
$\alpha_2=1/54 (-11+5 \sqrt{5})$. 
In the following we will obtain a normalized set of solutions of the Picard-Fuchs differential equation. From the index 
structure around the conifold  \ref{pfidx}, the existence of a logarithmic solution can be deduced. Furthermore, we have
solutions which start with $s^i$ ($s=(z-c)$, $i=0,1,2$) which we will denote by $\omega^c_i(s)$. We normalize the 
logarithmic solution $\log(s) \omega_1^c(s) + \cO(s^1)$ by requiring $\omega_1^c(s)=s+\cO(s^2)$. The solution corresponding
to the index $\rho_4=2$ is normalized to be of the form $\omega^c_2(s)=s^2 + \cO(s^3)$. A suitable linear combination with
$\omega_1^c(s)$ and $\omega_2^c(s)$ allows us to choose the solution for the index $\rho_1=0$ to be of the form

\begin{equation}
  \omega^c_0(s) = 1 + \cO(s^3).
\end{equation}

The mirror map can be now specified to be

\begin{equation}
  k_t t_c = \frac{\omega^c_1(s)}{\omega_0^c(s)},
\end{equation}
where $k_t$ is a constant which for the moment we can set to one. 

We solve the Picard-Fuchs equations over the ring $\mathbb{Q}[\alpha]/dis(\alpha)$ and obtain the following 
results for the periods and the mirror maps

\begin{eqnarray}
  \omega^{\alpha}_0(s)&=&1 + \frac{81}{250} (435709 + 1060776 \alpha) s^3 + \cO(s^4) \nonumber \\
  \omega^{\alpha}_1(s)&=& s - \frac{3}{50}(3709 + 9126 \alpha)s^2 + \frac{3}{25}(446957+1088046 \alpha )s^3 
                           + \cO(s^4) \nonumber \\ 
  \\
  s(t_{\alpha}) &=& t_{\alpha} -  \frac{3}{50}(3709 + 9126 \alpha) t_{\alpha}^2 
                    + \frac{3}{50} (770597 + 1875852 \alpha) t_{\alpha}^3 + \cO(t_{\alpha}^4)
\end{eqnarray}

In order to regain the solutions around the points $\alpha_i,~i\in\{1,2\}$ one has to substitute $\alpha$ by $\alpha_i$.
For more details about this method see \cite{HK}.

The topological limits around the conifold points are obtained by making the replacements

\begin{equation}
  A_1(s+c,\bar{s}+\bar{c}) \rightarrow (s+c) \frac{d}{ds}\log \frac{d t_c}{ds},~~
  B_k \rightarrow \frac{1}{\omega^c_0(s)}((s+c)\frac{d}{ds})^k \omega^c_0(s)
\end{equation}
in the defining relation \ref{gwp}.

The gap condition of \cite{HKQ} now tells us 

\begin{equation}
  \rF^{(g)}_c(t_c) = (\omega_0(s))^{2g-2} F_c^{(g)}(s)=\frac{\textrm{const.}}{t_c^{2g-2}} + \cO(t_c^0),
\end{equation}
for $g \geq  2$. This provides us with $(2g-2)-1$ equations which are vanishing conditions for the coefficients of 
$\frac{1}{t_c^i}(1 \leq i \leq 2g-3)$. Actually, the condition is even stronger as there exists a choice of the constant 
$k_t$ under which in all higher genus expansions the leading term is of the form 
$\frac{|B_{2g}|}{2g(2g-2)} \frac{1}{t_c^{2g-2}}$.

It is interesting to have a look at this gap structure in the expansions of Gromov-Witten potentials once the
holomorphic ambiguity is fixed completely,

\begin{eqnarray}
  \rF^{(2)}_{\alpha}(t_{\alpha}) &=& \frac{41-12276 \alpha}{874800 t_{\alpha}^2}
                                      + \frac{-14874743 + 3442099023 \alpha}{36450000} + O(t_{\alpha}), \nonumber \\
  \rF^{(3)}_{\alpha}(t_{\alpha}) &=& -\frac{5 (-15005 + 4493016 \alpha)}{4821232752 t_{\alpha}^4} + \cO(t_{\alpha}^0).
\end{eqnarray}

Again, substitute $\alpha$ by $\alpha_i$ to obtain the solutions around the specific vanishing point of the discriminant.

\subsubsection{Expansion around the orbifold point}

The index structure \ref{pfidx} of the Picard-Fuchs operator suggests that the point at infinity is a $\mathbb{Z}_3$ 
orbifold point. Therefore, we have to impose regularity of the free energies at this point in the moduli space. 
To obtain the topological limits we follow a path of argumentation presented in \cite{HK}. Let $x$ be the coordinate at 
infinity, i.e. $x=\frac{1}{z}$. Then we can define $\tilde{\cF}^{(g)}(x,\bar{x})$ to be the solutions
of the BCOV equation in $x$-coordinates with initial conditions $\tilde{\cF}^{(1)}_1(x,\bar{x})$ and 
$\tilde{\cF}^{(0)}_3=D_x D_x D_x \tilde{\cF}^{(0)}(x,\bar{x})$. 
On the other hand these initial conditions are related by

\begin{equation}
  \tilde{\cF}^{(0)}_3(x,\bar{x})=C_{xxx}(x)=C_{zzz}\Big(\frac{1}{x}\Big)\Big(\frac{dz}{dx}\Big)^3
                                =\cF^{(0)}_3\Big(\frac{1}{x},\frac{1}{\bar{x}}\Big)\Big(\frac{dz}{dx}\Big)^3.
\end{equation}

From this we can infer that $\tilde{\cF}^{(g)}(x,\bar{x})$ and $\cF^{(g)}(z,\bar{z})$ are in the same 
coordinate patch of a trivialization of the line bundle $\cL$, which again gives

\begin{equation}
  \tilde{\cF}^{(g)}(x,\bar{x}) = \cF^{(g)}\Big(\frac{1}{x},\frac{1}{\bar{x}}\Big).
\end{equation}

Therefore, the topological limit at infinity is simply obtained by setting 
$\tilde{\cF}^{(g)}(x,\bar{x})=\cF^{(g)}(A_1(\frac{1}{x},\frac{1}{\bar{x}}),
B_k(\frac{1}{x},\frac{1}{\bar{x}}),\frac{1}{x})$ and taking the limits

\begin{eqnarray}
  A_1\Big(\frac{1}{x},\frac{1}{\bar{x}}\Big) &=& \Big(\frac{dz}{dx} \frac{d\bar{z}}{d\bar{x}} G^{x\bar{x}}\Big)(-\theta_x)\Big(\frac{dx}{dz} 
                                          \frac{d\bar{x}}{d\bar{z}} G_{x\bar{x}}\Big) \rightarrow 
                                         -\Big(\frac{dx}{dt_{\infty}}\Big)\theta_x \Big(\frac{dt_{\infty}}{dx}\Big)-2 \\
  B_k\Big(\frac{1}{x},\frac{1}{\bar{x}}\Big) &=& e^{\tilde{K}(x,\bar{x})}(-\theta_x)^k e^{-\tilde{K}(x,\bar{x})} 
                                         \rightarrow \frac{1}{\omega^{\infty}_0(x)} (-\theta_x)^k \omega^{\infty}_0(x),
					 ~~(k=1,2,3),
\end{eqnarray}
where $\omega^{\infty}_0(x),\omega^{\infty}_1(x)$ and $t_{\infty}(x)=\frac{\omega^{\infty}_1(x)}{\omega^{\infty}_1(x)}$ 
are the periods and mirror map at infinity. 

So in order to proceed we have to calculate these quantities first. 
From the index structure we have the following set of solutions, $\omega^{\infty}_0(x)=x^{1/3} + \cO(x^{4/3})$,
$\omega^{\infty}_1(x)=x^{2/3} + \cO(x^{5/3})$, $\omega^{\infty}_2(x)=x^{4/3} + \cO(x^{7/3})$ and 
$\omega^{\infty}_3(x)=x^{5/3} + \cO(x^{8/3})$. Using a linear combination with $\omega^{\infty}_2(x)$ we can fix the 
first solution to be of the form

\begin{equation}
  \omega^{\infty}_0(x) = x^{1/3} + \cO(x^{7/3}).
\end{equation}

Furthermore, the second solution can be fixed by taking a linear combination with the third solution to

\begin{equation}
  \omega^{\infty}_1(x) = x^{2/3} + \cO(x^{8/3}).
\end{equation}

With these choices the relevant solutions are given by

\begin{eqnarray}
  \omega^{\infty}_0(x) &=& x^{1/3} + \frac{x^{7/3}}{131220} - \frac{67}{51018336} x^{10/3} + \cO(x^{13/3}), \nonumber \\
  \omega^{\infty}_1(x) &=& x^{2/3} - \frac{2}{45927} x^{8}{3} - \frac{467}{55801305} x^{11/3} + \cO(s^{14/3}), \nonumber \\
  x &=& t_{\infty}^3 - \frac{11}{102060} t_{\infty}^9 + \frac{12599}{595213920} t_{\infty}^12 + \cO(t_{\infty}^15).
\end{eqnarray}

Using these data and the holomorphic limit discussed above we obtain the following Gromov-Witten potentials

\begin{eqnarray}
  \rF^{(2)}_{\infty}(t_{\infty}) &=& \frac{\frac{41031}{160}+a_2}{t_{\infty}^4} 
                                   + \frac{\frac{1367}{80} + a_1}{t_{\infty}} + \cO(t_{\infty}), \nonumber \\
  \rF^{(3)}_{\infty}(t_{\infty}) &=& \frac{\frac{22453281}{1600}+a_4}{t_{\infty}^8} 
                                   + \frac{\frac{4572543}{3200} + a_3}{t_{\infty}^5}
                                   + \frac{-\frac{121464319}{567000} + a_2 + \frac{73 a_4}{229635}}{t_{\infty}^2}
                                   + \cO(t_{\infty}).
\end{eqnarray}

As the orbifold point is a conformal field theory point and thus has to be regular, we see that demanding the vanishing of the coefficients of
inverse powers of $t_{\infty}$ gives us $g$ conditions on the parameters of the holomorphic ambiguity.

Counting the number of boundary conditions from the orbifold and conifold points one notices that they are not yet enough to fix the ambiguity
completely. This is no problem for lower genera as the vanishing of lower degree Gopakumar-Vafa invariants gives us enough conditions to fix
all free parameters. On the other hand, as mentioned earlier, our example shows that there are not enough boundary conditions to solve the
model up to genus infinity.

\newpage

\section{Other Models}

We have analysed three other Calabi-Yau complete intersections in Grassmannians, namely $\left(\mathbb{G}(2,5)\right|\!\!\left| 1,2,2 \right)^{1}_{-120}$,
$\left(\mathbb{G}(3,6)\right|\!\!\left| 1^6 \right)^{1}_{-96}$ and $\left(\mathbb{G}(2,6)\right|\!\!\left| 1,1,1,1,2 \right)^{1}_{-116}$. 
All three admit interesting new features and share common properties with the model analysed previously. In particular, 
we have found a lense space point in the moduli space of the second model. 

\subsection{$\left(\mathbb{G}(2,5)\right|\!\!\left| 1,2,2 \right)^{1}_{-120}$}

The topological data of this Calabi-Yau are given by $\chi = -120$, $h^{2,1}=61$, $h^{1,1}=1$, $c_2 \cdot J=68$.
The Picard-Fuchs operator which was obtained in \cite{BFKS} admits the following index structure

\begin{center}

\begin{tabular}{c|cccc} \label{pfidx}
  z & 0 & $\alpha_1$ & $\alpha_2$ & $\infty$\\
  \hline \\
  $\rho_1$ & 0 & 0 & 0 & 1/2 \\
  $\rho_2$ & 0 & 1 & 1 & 1/2 \\
  $\rho_3$ & 0 & 1 & 1 & 3/2 \\
  $\rho_4$ & 0 & 2 & 2 & 3/2 \\
\end{tabular}

\end{center}

and the Yukawa coupling is determined to be 

\begin{equation}
  C_{zzz} = \frac{20}{z^3 (1-11 \cdot 2^4 z - 2^8 z^2)}.
\end{equation}

For the solutions around the conifold points we choose exactly the same normalization as in the case of 
$\left(\mathbb{G}(2,5)\right|\!\!\left| 1,1,3 \right)^{1}_{-150}$.
Looking at the point at infinity, we see that there are two logarithmic solutions. In order to obtain the mirror map 
only the first two solutions $\omega^{\infty}_0$ and $\omega^{\infty}_1$ are needed. They are of the form

\begin{eqnarray}
  \omega^{\infty}_0 & = & x^{1/2} + \cO(x^{5/2}), \nonumber \\
  \omega^{\infty}_1 & = & \log(x) x^{1/2} + \cO(x^{9/2}),
\end{eqnarray}
and we take the mirror map to be of the form $t = \frac{\omega^{\infty}_1(x)}{\omega^{\infty}_0(x)}$.

With these conventions we calculate the expansions of the free energies around the singular points of the moduli space.
We find the same gap conditions as in the case of $\left(\mathbb{G}(2,5)\right|\!\!\left| 1,1,3 \right)^{1}_{-150}$ 
around the two conifolds. The point at infinity turns out to be a regular point as we have to impose regularity on the 
Gromov-Witten potentials in order to obtain integral Gopakumar-Vafa numbers. We list the genus 2 and 3 expansions around this point

\begin{eqnarray}
  \rF^{2}_{\infty}(t_{\infty}) &=& \frac{5^{1/4} (136 + 3 a_2)}{48 \sqrt{3} t_{\infty}^{1/4}}
                                   + (a_1 + \frac{-119464 - 4047 a_2}{32000}) + \cO(t_{\infty}), \nonumber \\
  \rF^{3}_{\infty}(t_{\infty}) &=& \frac{\sqrt{5} (\frac{1024}{3} + a_4)}{768 \sqrt{t_{\infty}}}
                                   + \frac{-28849664 + 144000 a_3 - 36423 a_4}{460800 \sqrt{3} 5^{3/4} t_{\infty}^{1/4}}
                                   + \cO(t_{\infty}).
\end{eqnarray}

As one can see regularity restrictions give us $g-1$ boundary conditions on the ambiguity.

\subsection{$\left(\mathbb{G}(3,6)\right|\!\!\left| 1^6 \right)^{1}_{-96}$}

This Calabi-Yau has the topological data $\chi = -96$, $h^{2,1}=49$,$h^{1,1}=1$, $c_2 \cdot J=84$.
The Picard-Fuchs operator given in \cite{BFKS}  admits the following index structure

\begin{center}

\begin{tabular}{c|cccc} \label{pfidx}
  z & 0 & $\alpha_1$ & $\alpha_2$ & $\infty$\\
  \hline \\
  $\rho_1$ & 0 & 0 & 0 & 4/3 \\
  $\rho_2$ & 0 & 1 & 1 & 1 \\
  $\rho_3$ & 0 & 1 & 1 & 1 \\
  $\rho_4$ & 0 & 2 & 2 & 5/4 \\
\end{tabular}

\end{center}

The Yukawa coupling is given by

\begin{equation}
  C_{zzz} = \frac{28}{z^3 (1- 26 \cdot 2^2 z - 27 \cdot 2^4 z^2)}.
\end{equation}

The point at infinity admits one logarithmic solution which corresponds to a vanishing cycle and it appears that it 
also admits some orbifold features. The mirror map is given by $t = \frac{\omega^{\infty}_1(x)}{\omega^{\infty}_0(x)}$,
where

\begin{eqnarray}
  \omega^{\infty}_0 & = & x^{3/4} + \cO(x^{7/4}), \nonumber \\
  \omega^{\infty}_1 & = & x + \cO(x^2).
\end{eqnarray}

An interesting feature of this model is the fact that the two vanishing points of the discriminant, although having
the same Picard-Fuchs-indices, behave differently when we analyze the Gromov-Witten potentials. In particular, the
genus 1 Gromov-Witten potential of this model is

\begin{equation}
  \rF^{(1)}(z) = \frac{1}{2} \log \left\{\left(\frac{1}{\omega_0(z)}\right)^{3+h^{1,1}
               -\frac{\chi}{12}} \left(\frac{dz}{dt}\right) (-1+z)^{-\frac{1}{3}} (-1+64 z)^{-\frac{1}{6}}
                z^{-1-\frac{c2\cdot H}{12}}\right\}.
\end{equation}

This suggests that the point $z=1$ is not an ordinary conifold point but rather a lense space point, that is a point, 
where a cycle $\mathcal{C}$(for example $S^3$) modded by a group $G$ shrinks to zero size. 
In the case of $\mathcal{C}=S^3$ $G$ is a discrete subgroup of $SU(2)$ and the resulting space $S^3/G$ has fundamental 
group $G$. 
Spaces of this form where investigated in \cite{GV}, where the number of BPS states admitted 
by such cycles was calculated. There it was argued that the number of D-brane bound states which are
BPS is equal to the number of irreducible representations of $G$ and their mass is given by the formula 
$M_i = \mu d_i / G$ where $\mu$ is the size of the unmodded cycle and $d_i$ is the dimension of the $i$th irreducible 
representation of $G$. Comparing this with the genus one free energy of the topological string one finds

\begin{equation}
  F^{(1)} = \sum_i - \frac{1}{12} \log(M_i) = \sum_i -\frac{1}{12} \log(\mu d_i / G).
\end{equation} 

In our particular example this is

\begin{equation}
  F^{(1)} = -\frac{1}{12} \log(t_{1/64}) - \frac{2}{12} \log(t_1).
\end{equation}

Using the identification $t_1 = \mu / 2$ we find from the above formula that the group $G$ must be $\mathbb{Z}_2$.
This also shows that two hypermultiplets are becoming massless at $z=1$.

Our result is supported by the monodromy calculations made in \cite{ES}. There it was found that the monodromy matrix at the
point $z=1$ is of Picard-Lefschetz form $S_{\lambda,v}$, where $\lambda=2$ which shows that this point is not an
ordinary conifold point. 

Higher genus calculations show that the ordinary gap condition holds at $z=1/64$ which is to be expected as this point
is a conifold point. On the other hand the gap condition has to be slightly modified around $z=1$. If we assume that
the two hypermultiplets becoming massless are not interacting the modification to the leading term of the higher genus
Gromov-Witten potential reads as follows

\begin{equation}
  \rF^{g}_1(t_1) = 2 \frac{|B_{2g}|}{2g(2g-2)} \frac{1}{\mu^{2g-2}} + \cO(t_1^0) 
                 = 2 \frac{|B_{2g}|}{2g(2g-2)} \frac{1}{2^{2g-2}} \frac{1}{t_1^{2g-2}} + \cO(t_1^0).
\end{equation}

This is exactly what we observe.

It remains to be discussed the point at infinity. It admits a gap-like structure as can be seen for example from
the genus 4 expansion

\begin{eqnarray}
  \rF^{4}_{\infty}(t_{\infty}) &=& \frac{7}{240~ t_{\infty}^6} + \frac{101797151}{11010048000} t_{\infty}^2 + \cO (t_{\infty}^3).
\end{eqnarray}

\subsection{$\left(\mathbb{G}(2,6)\right|\!\!\left| 1,1,1,1,2 \right)^{1}_{-116}$}

This manifold is characterized by the data $\chi = -116$, $h^{2,1}=59$, $h^{1,1}=1$, $c_2 \cdot J=76$.
The structure of the solutions of the Picard-Fuchs operator is the following

\begin{center}

\begin{tabular}{c|cccc} \label{pfidx}
  z & 0 & $\alpha_1$ & $\alpha_2$ & $\infty$\\
  \hline \\
  $\rho_1$ & 0 & 0 & 0 & 1/2 \\
  $\rho_2$ & 0 & 1 & 1 & 2/3 \\
  $\rho_3$ & 0 & 1 & 1 & 4/3 \\
  $\rho_4$ & 0 & 2 & 2 & 3/2 \\
\end{tabular}

\end{center}

The Yukawa coupling is given by

\begin{equation}
  C_{zzz} = \frac{42}{z^3 (1 - 65 z - 64 z^2)}.
\end{equation}

The conifold locus is treated as usual. The mirror map at $z=\infty$ is obtained by taking the ratio of the first two 
periods. They are of the form

\begin{eqnarray}
  \omega^{\infty}_0 & = & x^{1/2} + \cO(x^{5/2}), \nonumber \\
  \omega^{\infty}_1 & = & x^{2/3} + \cO(x^{5/3}).
\end{eqnarray}

Now, our calculations show that the gap condition holds at the conifold locus. Furthermore, the point at infinity 
at first sight seems to be a regular orbifold point with $\mathbb{Z}_6$-symmetry and indeed this seems to be the case
up to genus 3. But at genus 4 we find that the expansion of the Gromov-Witten potential around this point is singular.
In particular we find 

\begin{eqnarray}
  \rF^{4}_{\infty}(t_{\infty}) & = & \frac{-\frac{8606402923}{164640} + a_6}{t_{\infty}^{18}} 
                                      + \frac{-\frac{500305024099}{49787136} + a_5 - \frac{10}{63} a_6}{t_{\infty}^{12}} \nonumber \\
                               & ~ &  + \frac{-\frac{443407050538901893}{179412923289600} + a_4 - \frac{20}{189} a_5 
                                      + \frac{831575}{54486432} a_6}{t_{\infty}^6} + \cO(t_{\infty}^0),
\end{eqnarray}
before fixing the ambiguity and 

\begin{equation}
  \rF^{4}_{\infty}(t_{\infty}) = \frac{2}{2187~t_{\infty}^6} + \frac{108172361}{131681894400} + \cO(t_{\infty}),
\end{equation}
after having fixed the ambiguity.

\subsection{$\left(\mathbb{G}(2,7)\right|\!\!\left| 1^7 \right)^{1}_{-98}$}

This manifold is characterized by the data $\chi = -98$, $h^{2,1}=50$, $h^{1,1}=1$, $c_2 \cdot J=84$.
The structure of the solutions of the Picard-Fuchs operator is the following

\begin{center}

\begin{tabular}{c|cccccc} \label{pfidx}
  z & 0 & $\alpha_1$ & $\alpha_2$ & $\alpha_3$ & $3$ & $\infty$\\
  \hline \\
  $\rho_1$ & 0 & 0 & 0 & 0 & 0 & 1\\
  $\rho_2$ & 0 & 1 & 1 & 1 & 1 & 1\\
  $\rho_3$ & 0 & 1 & 1 & 1 & 3 & 1\\
  $\rho_4$ & 0 & 2 & 2 & 2 & 4 & 1\\
\end{tabular}

\end{center}

We see that the Picard-Fuchs differential operator has the property of maximally degeneration at both $z=0$ and $z=\infty$.
It was found in \cite{Ro} that the expansion about $z=0$ corresponds to the K\"ahler moduli of the Grassmannian Calabi-Yau 
$M = \left(\mathbb{G}(2,7)\right|\!\!\left| 1^7 \right)^{1}_{-98}$, and the expansion about $z=\infty$ to that of a Pfaffian 
Calabi-Yau $M'$. In \cite{HK} the instanton calculations for this model were extended up to genus 5 and we confirm their 
results for low genus.

\section{Conclusions}

In this paper we analyzed the topological string on five one parameter
Calabi-Yau spaces realized as complete intersections in Grassmannians. 
One result is that the gap condition at the conifold that was 
discovered in~local geometries in~\cite{Huang:2006si} and  global geometries 
in~\cite{HKQ} is also present in the Grassmannian Calabi-Yau manifolds.

Since it involves subleading terms the gap condition is more then a 
local statement. The fact that leading behavior of the  $\rF_g(t_c)$ near 
the conifold point is given by the $c=1$ string is understood from the 
leading order local geometry of the nodal singularity~
\cite{Ghoshal:1995wm,Dijkgraaf:2003xk,ADKMV} and is true in any 
choice of the local coordinate system which has the right scaling behavior 
of the complex structure parameterization. On the other hand the gap is 
sensitive to the global embedding, because it is only true in the flat 
coordinates for the complex structure parameters, whose form depends on 
global properties of the period integrals. 

Unlike the toric one parameter Calabi-Yau the Grassmannian one parameter models 
have usually several conifolds at various values of $z$ in their moduli space 
and all these have to fulfill the gap condition in order for the BPS invariants to be 
integer. In all cases we found explicitly integer  BPS numbers for the
symplectic invariants up to genus 5, which would be very interesting 
to confirm by methods of enumerative geometry.  

We find  that the model  $\left(\mathbb{G}(3,6)\right|\!\!\left| 1^6 \right)^{1}_{-96}$ 
has a conifold at $z=\frac{1}{64}$ and a lense space $S^3/\mathbb{Z}_2$ shrinking at $z=1$. 
We find that at the lense space singularity the analysis of the leading terms 
is exactly as predicted in~\cite{GV} and that in addition there is a full
gap structure in the subleading terms. The physical interpretation is that 
the two BPS states do not interact and in particular do not form light 
bound states. This model has also at $t_\infty$ a branch point of
order $12$, a  single logarithmic solution and a full gap structure.  

The models $\left(\mathbb{G}(2,5)\right|\!\!\left| 1,1,3 \right)^{1}_{-150}$,
$\left(\mathbb{G}(2,5)\right|\!\!\left| 1,2,2 \right)^{1}_{-120}$ are 
regular at $t_\infty=0$ at least to genus 5. The first has 
regular solutions, which hints a CFT with an $\mathbb{Z}_3$ automorphism 
at $t_\infty=0$. In this model the BPS invariant  $n_6^4=5$ has  
been checked geometrically by Sheldon Katz, who found also the vanishing
of the BPS invariants for the other model in accord with 
Castelnouvos Theory.   

The model $\left(\mathbb{G}(2,5)\right|\!\!\left| 
1,2,2 \right)^{1}_{-120}$ has two logarithmic solutions and a 
branch point order of $2$. It is conceivable that higher $\rF_g$
are not regular at  $t_\infty=0$.

The model $\left(\mathbb{G}(2,6)\right|\!\!\left| 1,1,1,1,2
\right)^{1}_{-116}$ has two different conifolds with a full 
gap structure. At the point $t_\infty=0$ it has regular 
solutions with an $\mathbb{Z}_6$ branching. Curiously we find 
that the integrality of the BPS require that it has 
singular behavior in the $\rF_g$ for $g>3$.

For the Rodland example $\left(\mathbb{G}(2,7)\right|\!\!\left| 1^7
\right)^{1}_{-98}$, which has two points of 
maximal unipotent monodromy  we confirm the analysis 
of~\cite{HK} for low genus.  

Solving the topological string 
to all genus would be important to study black holes in five 
and four dimensions~\cite{Huang:2007sb}. It is notable that the 
range of the topological data, which determine the 
semiclassical analysis of black holes take more extreme 
values for the Grassmannians than for the toric varieties.
In particular $c_2\cdot H$ and the triple intersection 
$H^3$ take the highest values for Grassmannian Calabi-Yau. 
This is very useful for comparing the semiclassical and 
the microscopic description of black holes along the 
lines of~\cite{Huang:2007sb}. Indeed we find that the 
microscopic entropy the Richardson transforms converge
within $4$ \% to the expected value of the macroscopic 
calculation. For reference we show one plot for the
extreme value of $H^3=42$ in Appendix C.

\section*{Acknowledgments}
We would like to thank Thomas Grimm, Sheldon Katz, Marcos Mari\~no, 
Min-xin Huang, Piotr Sulkowski, and Don Zagier for very  valuable discussions.

\bigskip \bigskip

\appendix

\noindent {\bf \Large Appendix}

\section{Chern classes and topological invariants}

\begin{tabular}{lll}
  \multicolumn{3}{c}{~}\\ 
  $ \mathbb{G}(2,5)$         & : & $\int_{G(2,5)} \sigma_1^6 = 5, ~ \int_{G(2,5)} \sigma_2 \sigma_1^4 = 3,
                                                      ~ \int_{G(2,5)} \sigma_3 \sigma_1^3=1$,\\
  \multicolumn{3}{c}{~}\\ 
  $\left(\mathbb{G}(2,5)\right|\!\!\left| 1,1,3 \right)^{1}_{-150}$  & : & $c(\left(\mathbb{G}(2,5)\right|\!\!\left| 1,1,3 \right)^{1}_{-150})$\\  
  ~                                                  & ~ & $= 1 + (5 c_1(Q)^2 - c_2(Q))$\\ 
  ~                                                  & ~ & $~- (8 c_1(Q)^3 + 5 c_1(Q) c_2(Q) - 5 c_3(Q)) + \cdots$, \\
  \multicolumn{3}{c}{~}\\ 
  \multicolumn{3}{c}{ $\Rightarrow$ $\chi=-150$,~~$c_2 \cdot H = 66$,~~$H^3 = 15$.}\\
  \multicolumn{3}{c}{~}\\
  $\left(\mathbb{G}(2,5)\right|\!\!\left| 1,2,2 \right)^{1}_{-120}$  & : & $c(\left(\mathbb{G}(2,5)\right|\!\!\left| 1,2,2 \right)^{1}_{-120})$\\  
  ~                                                  & ~ & $= 1 + (4 c_1(Q)^2 - c_2(Q))$ \\
  ~                                                  & ~ & $~- (4 c_1(Q)^3 + 5 c_1(Q) c_2(Q) - 5 c_3(Q)) + \cdots$, \\
  \multicolumn{3}{c}{~}\\ 
  \multicolumn{3}{c}{ $\Rightarrow$ $\chi=-120$,~~$c_2 \cdot H = 68$,~~$H^3=20$.}\\
  \multicolumn{3}{c}{~}\\
  $ \mathbb{G}(2,6)$         & : & $\int_{G(2,6)} \sigma_1^8 = 14$, ~ $\int_{G(2,6)} \sigma_2 \sigma_1^6=9$,
                                                            ~$\int_{G(2,6)} \sigma_3 \sigma_1^5 = 4$,\\
  \multicolumn{3}{c}{~}\\
  $\left(\mathbb{G}(2,6)\right|\!\!\left| 1,1,1,1,2 \right)^{1}_{-116}$ & : & $c(\left(\mathbb{G}(2,6)\right|\!\!\left| 1,1,1,1,2 \right)^{1}_{-116})$\\    
  ~                                                   & ~ & $ = 1 + (4 c_1(Q)^2 - 2 c_2(Q))$\\
  ~                                                   & ~ & $ ~ - (2 c_1(Q)^3 + 6 c_1(Q) c_2(Q) - 6 c_3(Q)) + \cdots$,\\
  \multicolumn{3}{c}{~}\\
  \multicolumn{3}{c}{ $\Rightarrow$ $\chi=-116$,~~$c_2 \cdot H = 76$,~~$H^3 = 28$.}\\
  \multicolumn{3}{c}{~}\\
  $ \mathbb{G}(3,6)$         & : & $\int_{G(3,6)} \sigma_1^9 =42$, ~ $\int_{G(3,6)} \sigma_2 \sigma_1^7 =21$,
                                                            ~$\int_{G(3,6)} \sigma_3 \sigma_1^6 = 5$,\\
  \multicolumn{3}{c}{~}\\
  $\left(\mathbb{G}(3,6)\right|\!\!\left| 1^6 \right)^{1}_{-96}$     & : & $c(\left(\mathbb{G}(3,6)\right|\!\!\left| 1^6 \right)^{1}_{-96})$ \\
  ~                                                   & ~ & $= 1 + 2 c_1(Q)^2$\\
  ~                                                   & ~ & $~ - (6 c_1(Q) c_2(Q) - 6 c_3(Q))+\cdots$,\\
  \multicolumn{3}{c}{~}\\
  \multicolumn{3}{c}{ $\Rightarrow$ $\chi=-96$,~~$c_2 \cdot H = 84$,~~$H^3=42$.}\\
  \multicolumn{3}{c}{~}\\
  $ \mathbb{G}(2,7)$         & : & $\int_{G(2,7)} \sigma_1^{10} =42$, ~ $\int_{G(2,7)} \sigma_2 \sigma_1^8 =28$,
                                                            ~$\int_{G(2,7)} \sigma_3 \sigma_1^7 = 14$,\\
  \multicolumn{3}{c}{~}\\
  $\left(\mathbb{G}(2,7)\right|\!\!\left| 1^7 \right)^{1}_{-98}$ & : & $c(\left(\mathbb{G}(2,7)\right|\!\!\left| 1^7 \right)^{1}_{-98})$\\ 
  ~                                                   & ~ & $= 1 + (4 c_1(Q)^2 - 3 c_2(Q))$\\
  ~                                                   & ~ & $~ - (7 c_1(Q) c_2(Q)- 7 c_3(Q)) + \cdots$, \\
  \multicolumn{3}{c}{~}\\
  \multicolumn{3}{c}{ $\Rightarrow$ $\chi=-98$,~~$c_2 \cdot H = 84$,~~$H^3=42$.}\\
\end{tabular}

\section{Tables of Gopakumar-Vafa invariants}

\begin{landscape}
\begin{table}[t]
\begin{center}
\begin{tabular}{c|llllll}
  d  &         $g=0$         &         $g=1$         &          $g=2$           & $g=3$                &  $g=4$            & $g=5$\\
  \hline \\
   1 & 540                   & 0                     & 0                        & 0                    & 0                 & 0          \\
   2 & 12555                 & 0                     & 0                        & 0                    & 0                 & 0          \\
   3 & 621315                & -1                    & 0                        & 0                    & 0                 & 0          \\
   4 & 44892765              & 13095                 & 0                        & 0                    & 0                 & 0          \\
   5 & 3995437590            & 17230617              & -1080                    & 0                    & 0                 & 0          \\
   6 & 406684089360          & 6648808835            & 921735                   & 420                  & 5                 & 0          \\
   7 & 45426958360155        & 1831575868830         & 6512362740               & -26460               & -2160             & 0          \\
   8 & 5432556927598425      & 433375127634753       & 5837267557035            & 6528493485           & 218160            & -2160      \\
   9 & 684486974574277695    & 94416986839804040     & 3061620003073095         & 20216637579465       & 6735865790        & 2770635    \\
  10 & 89872619976165978675  & 19571240651198871015  & 1223886411726167880      & 22818718255545315    & 85314971897190    & 5441786955 \\
\end{tabular}
\end{center}
\caption{Gopakumar-Vafa invariants $n_g(d) (g \leq 5)$ of the Grassmannian Calabi-Yau threefold 
$\left(\mathbb{G}(2,5)\right|\!\!\left| 1,1,3 \right)^{1}_{-150}$.}\label{table1}
\end{table}

\begin{table}[t]
\begin{center}
\begin{tabular}{c|llllll}
  d  &         $g=0$         &         $g=1$         &          $g=2$           & $g=3$                &  $g=4$            & $g=5$ \\
  \hline \\
   1 & 400                   & 0                     & 0                        & 0                    & 0                 & 0     \\
   2 & 5540                  & 0                     & 0                        & 0                    & 0                 & 0     \\
   3 & 164400                & 0                     & 0                        & 0                    & 0                 & 0     \\
   4 & 7059880               & 1537                  & 0                        & 0                    & 0                 & 0     \\
   5 & 373030720             & 882496                & 0                        & 0                    & 0                 & 0     \\
   6 & 22532353740           & 214941640             & 15140                    & 0                    & 0                 & 0     \\
   7 & 1493352046000         & 37001766880           & 57840400                 & -800                 & 0                 & 0     \\
   8 & 105953648564840       & 5388182343297         & 36620960080              & 10792630             & 320               & 5     \\
   9 & 7919932042500000      & 715201587952800       & 12817600017680           & 33952864320          & 697600            & -1600 \\
  10 & 616905355407694800    & 89732472170109248     & 3295335805457360         & 29386059424200       & 32052405340       & -32320\\
\end{tabular}
\end{center}
\caption{Gopakumar-Vafa invariants $n_g(d) (g \leq 5)$ of the Grassmannian Calabi-Yau threefold 
$\left(\mathbb{G}(2,5)\right|\!\!\left| 1,2,2 \right)^{1}_{-120}$.}
\end{table}
\end{landscape}

\newpage

\begin{landscape}
\begin{table}[t]
\begin{center}
\begin{tabular}{c|llllll}
  d  &         $g=0$         &         $g=1$         &          $g=2$           & $g=3$                &  $g=4$      & $g=5$         \\
  \hline \\
   1 & 210                   & 0                     & 0                        & 0                    & 0           & 0\\
   2 & 1176                  & 0                     & 0                        & 0                    & 0           & 0\\
   3 & 13104                 & 0                     & 0                        & 0                    & 0           & 0\\
   4 & 201936                & 0                     & 0                        & 0                    & 0           & 0\\
   5 & 3824016               & 84                    & 0                        & 0                    & 0           & 0\\
   6 & 82568136              & 74382                 & 0                        & 0                    & 0           & 0\\
   7 & 1954684008            & 8161452               & 0                        & 0                    & 0           & 0\\
   8 & 49516091520           & 560512344             & 70896                    & 0                    & 0           & 0\\
   9 & 1321186053432         & 31354814820           & 39198978                 & 0                    & 0           & 0\\
  10 & 36729091812168        & 1568818990200         & 7239273552               & 1086246              & 0           & 0\\
  11 & 1055613263065704      & 73339159104540        & 827701960638             & 932836632            & 1722        & 0\\
  12 & 31184875579315920     & 3279169536538154      & 72679697259288           & 284870410986         & 55653752    & 0\\
\end{tabular}
\end{center}
\caption{Gopakumar-Vafa invariants $n_g(d) (g \leq 5)$ of the Grassmannian Calabi-Yau threefold 
$\left(\mathbb{G}(3,6)\right|\!\!\left| 1^6 \right)^{1}_{-96}$.}\label{table1}
\end{table}

\begin{table}[t]
\begin{center}
\begin{tabular}{c|llllll}
  d  &         $g=0$         &         $g=1$         &          $g=2$           & $g=3$                &  $g=4$          & $g=5$\\
  \hline \\
   1 & 280                   & 0                     & 0                        & 0                    & 0               & 0\\
   2 & 2674                  & 0                     & 0                        & 0                    & 0               & 0\\
   3 & 48272                 & 0                     & 0                        & 0                    & 0               & 0\\
   4 & 1279040               & 27                    & 0                        & 0                    & 0               & 0\\
   5 & 41389992              & 26208                 & 0                        & 0                    & 0               & 0\\
   6 & 1531603276            & 5914124               & -54                      & 0                    & 0               & 0\\
   7 & 62153423432           & 745052912             & 56112                    & 0                    & 0               & 0\\
   8 & 2699769672096         & 73219520613           & 120462612                & -5267                & 0               & 0\\
   9 & 123536738915800       & 6326648922384         & 40927354944              & 4713072              & 840             & 0\\
  10 & 5890247824324990      & 506932941439940       & 8145450103430            & 15699104736          & -91464          & -404\\
  11 & 290364442225572848    & 38717395881042032     & 1228133118935408         & 8307363701728        & 4174512664      & 66640\\
  12 & 14713407331980050400  & 2863231551878100494   & 156147718274297768       & 2460694451990694     & 7534787308968   & 991403118\\
\end{tabular}
\end{center}
\caption{Gopakumar-Vafa invariants $n_g(d) (g \leq 5)$ of the Grassmannian Calabi-Yau threefold 
$\left(\mathbb{G}(2,6)\right|\!\!\left| 1,1,1,1,2 \right)^{1}_{-116}$.}
\end{table}

\begin{table}[t]
\begin{center}
\begin{tabular}{c|llllll}
  d  &         $g=0$         &         $g=1$         &          $g=2$           & $g=3$                &  $g=4$          & $g=5$\\
  \hline \\
   1 & 196                   & 0                     & 0                        & 0                    & 0               & 0\\
   2 & 1225                  & 0                     & 0                        & 0                    & 0               & 0\\
   3 & 12740                 & 0                     & 0                        & 0                    & 0               & 0\\
   4 & 198058                & 0                     & 0                        & 0                    & 0               & 0\\
   5 & 3716944               & 588                   & 0                        & 0                    & 0               & 0\\
   6 & 79823205              & 99960                 & 0                        & 0                    & 0               & 0\\
   7 & 1877972628            & 8964372               & 0                        & 0                    & 0               & 0\\
   8 & 47288943912           & 577298253             & 99960                    & 0                    & 0               & 0\\
   9 & 1254186001124         & 31299964612           & 47151720                 & -1176                & 0               & 0\\
  10 & 34657942457488        & 1535808070650         & 7906245550               & 325409               & 0               & 0\\
  11 & 990133717028596       & 70785403788680        & 858740761340             & 956485684            & -25480          & 3675\\
  12 & 29075817464070412     & 3129139504135680      & 73056658523632           & 301227323110         & 27885116        & 73892\\
\end{tabular}
\end{center}
\caption{Gopakumar-Vafa invariants $n_g(d) (g \leq 5)$ of the Grassmannian Calabi-Yau threefold 
$\left(\mathbb{G}(2,7)\right|\!\!\left| 1^7 \right)^{1}_{-98}$.}
\end{table}

\end{landscape}

\newpage

\begin{landscape}

\begin{table}[t]
\begin{center}
\begin{tabular}{c|llll}
  d  &         $g=0$             &         $g=1$            &          $g=2$           & $g=3$                   \\
  \hline \\
   1 & 588                       & 0                        & 0                        & 0                       \\
   2 & 12103                     & 0                        & 0                        & 0                       \\
   3 & 583884                    & 196                      & 0                        & 0                       \\
   4 & 41359136                  & 99960                    & 0                        & 0                       \\
   5 & 3609394096                & 34149668                 & 12740                    & 0                       \\
   6 & 360339083307              & 9220666238               & 25275866                 & 1225                    \\
   7 & 39487258327356            & 2163937552736            & 21087112172              & 22409856                \\
   8 & 4633258198646014          & 466455116030169          & 11246111235996           & 58503447590             \\
   9 & 572819822939575596        & 95353089205907736        & 4601004859770928         & 67779027822044          \\
  10 & 73802503401477453288      & 18829753458134112872     & 1586777390750641117      & 50069281882780727       \\
\end{tabular}
\end{center}
\end{table}

\begin{table}[t]
\begin{center}
\begin{tabular}{c|ll}
  d  &         $g=4$             & $g=5$\\
  \hline \\
   1 & 0                         & 0\\
   2 & 0                         & 0\\
   3 & 0                         & 0\\
   4 & 0                         & 0\\
   5 & 0                         & 0\\
   6 & 0                         & 0\\
   7 & 0                         & 0\\
   8 & 25371416                  & 3675\\
   9 & 216888021056              & 33575388\\
  10 & 521484626374894           & 1111788286385\\
\end{tabular}
\end{center}
\caption{Gopakumar-Vafa invariants $n_g(d) (g \leq 5)$ of the Pfaffian Calabi-Yau threefold $M'$.}
\end{table}

\end{landscape}

\newpage

\section{5D Blackhole asymptotic}
\begin{figure}[ht]
\begin{center}
\epsfig{file=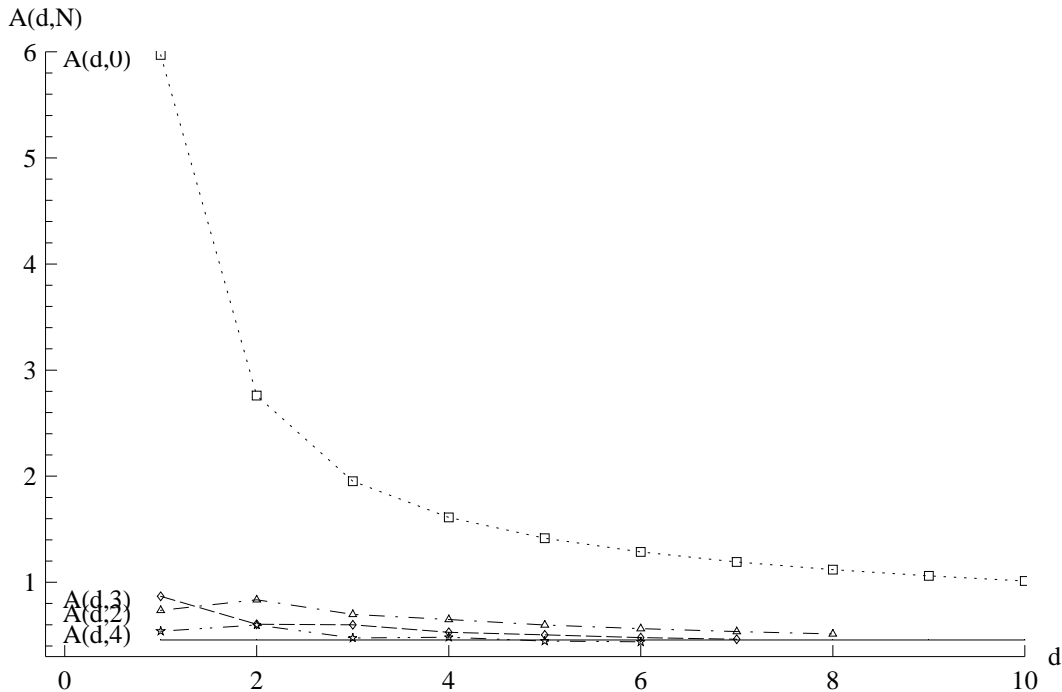, width=14cm}
\end{center}
\caption{Leading behavior of the microscopic entropy for the 5d black
  hole for the Grassmannian Calabi-Yau threefold
  $\left(\mathbb{G}(2,7)\right|\!\!\left| 1,1,1,1,1,1,1
  \right)^{1}_{-98}$. $A(d,m)$ are the Richardson transforms.
 The Richardson transforms of the microscopic entropy converge within 
$4$ \% to the expected value from the macroscopic calculation $b_0=\frac{4
  \pi }{3 \sqrt{2 H^3}}\sim .046$ for $H^3=42$, see \cite{Huang:2007sb} for details.} 
\label{leading}
\end{figure}

\def\plb#1 #2 {Phys. Lett. {\bf B#1} #2 }
\def\npb#1 #2 {Nucl. Phys. {\bf B#1} #2 }
\def\cmp#1 #2 {Commun. Math. Phys. {\bf #1} #2 }
\def\cqg#1 #2 {Class.Quant.Grav. {\bf #1} #2 }
\def\jgp#1 #2 {J. Geom. Phys. {\bf #1} #2 }
\def\atmp#1 #2 {Adv. Theor. Math. Phys. {\bf #1} #2 }


\end{document}